\documentclass[%
preprint,
amsmath,amssymb,
prc,
]{revtex4-2}

\usepackage[dvipdfmx]{graphicx}
\usepackage{dcolumn}
\usepackage{bm}
\usepackage{braket}
\usepackage{color}

\begin{document}
\preprint{}

\title{Continuum random-phase approximation for gamma transition between excited  states in neutron-rich nuclei}

\author{Teruyuki Saito}
 \affiliation{Graduate School of Science and Technology, Niigata University, Niigata 950-2181, Japan}
\author{Masayuki Matsuo}
\affiliation{Department of Physics, Faculty of Science. Niigata University, Niigata 950-2181, Japan}

\date{May 17, 2021}

\begin{abstract}
A characteristic feature of collective and particle-hole excitations in neutron-rich nuclei
is that many of them couple to unbound neutron in continuum single-particle orbits.
The continuum random phase approximation (cRPA) 
is a powerful many-body method that describes such excitations, and
it provides a scheme to evaluate transition strengths from the ground state. 
In an attempt to apply cRPA to the
 radiative neutron capture reaction, we 
 formulate in the present study an extended scheme of cRPA that describes 
 gamma-transitions from the excited states under consideration, which decay 
 to  low-lying excited states as well as the ground state.  
 This is achieved by introducing a non-local one-body operator
 which causes transitions to a low-lying excited state, and describing a
 density-matrix response against this operator. 
As a demonstration of this new scheme, we perform numerical calculation for
dipole, quadrupole, and octupole excitations in $^{140}$Sn, and discuss
E1 and E2  transitions decaying to low-lying $2^{+}_{1,2}$ and $3^{-}_{1}$
states. The results point to cases where the 
branching ratio to the low-lying states is larger than or comparable with that to the ground state. We
 discuss key roles of collectivity and continuum  orbits in both initial and final states.

\end{abstract}

\maketitle


\section{Introduction}

Theoretical and experimental studies  of neutron-rich nuclei have been performed extensively in recent years, and they revealed peculiar features which are
related to the small neutron separation energy or the weak binding of the last neutrons. Examples include 
the pygmy dipole resonance or the soft dipole excitation, which are considered as a new type of collective excitation or a continuum particle-hole 
excitations where a neutron is brought into unbound scattering state~\cite{Hansen1987, Suzuki1990, Bertsch1991, Paar2007}.  In the laboratory experiments, such exotic modes of excitation are observed 
in the excitation reactions such as the photo-absorption or the Coulomb or nuclear dissociation processes~\cite{Savran2013, Aumann2019, Aumann2013}. In the nature, neutron-rich
nuclei play important role in the r-process nucleosynthesis, and it has been pointed out that the pygmy or the soft dipole excitations
might influence the radiative neutron-capture reaction and resultant abundance of r-process nuclei~\cite{Goriely1998, Arnould2007}.

The radiative neutron-capture reaction are usually considered in terms of two different mechanisms:
the statistical or compound nuclear (CN) process and the direct capture (DC) process~\cite{Arnould2007}. The CN process is dominant in
nuclei with relatively large neutron separation energy and high level density 
while the DC process becomes dominant in nuclei close to neutron-drip line~\cite{Arnould2007, Mathews1983, Goriely1997}. 
For the CN process, the Hauser-Feshbach statistical model is assumed and
the role of the exotic excitation modes are usually taken into account via the gamma-decay strength
function~\cite{Arnould2007}. For the DC process, however, the exotic modes need to be
described explicitly as a doorway state of the neutron capture, and also do  gamma-decays from the 
populated excited state. Such theoretical descriptions of the DC process, applicable to
the medium and heavy neutron-rich nuclei (relevant to the r-process),  have not been formulated, in our knowledge,
except our preceding study\cite{Matsuo2015}. The DC models applied so far to medium and heavy nuclei adopt the independent particle model~\cite{Lane1960, Raman1985, Mengoni1995, Rauscher1998, Rauscher2010, Xu2012}, in which the 
collective effect is neglected.
The semi-direct model\cite{Bonneau2007,Chiba2008} takes into account the effect of the giant resonance is proposed, but it
is essentially the same as the independent particle model as far as the r-process neutron-capture at very low neutron energy
is concerned.

In the previous publication~\cite{Matsuo2015}, we adopted the continuum quasiparticle random-phase approximation (cQRPA) based on the 
density functional theory to describe the DC process via the exotic excitation modes. We describe the Coulomb
excitation or photo-absorption of an even-even neutron-rich
nucleus $A$, leading to an excited state ${\rm A^{*} \to (A-1) +n}$ which may emit a neutron if the excitation energy exceeds
the neutron threshold. Collective
correlations are taken into account in the linear response framework
 to calculate the strength function, and the Green's function method ~\cite{Bertsch1975, Shlomo1975, Matsuo2001} enables us to include 
 the unbound single-neutron state with a scattering wave.
By decomposing the strength function  into different channels of ${\rm (A-1) + n}$ with
 a method of
Zangwill and Soven~\cite{Zangwill1980} and using the
reciprocity theorem, we obtain the cross section of the radiative direct neutron-capture 
${\rm (A - 1) + n \to A^{*} \to A+ \gamma}$.  We remark however that further improvement is needed in this
approach since the gamma transitions ${\rm A^{*} \to A^{**}+ \gamma}$ decaying to 
low-lying excited states $A^{**}$  also occur in reality.

In the present study and in subsequent papers, we intend to extend the approach of Ref.~\cite{Matsuo2015}
in order to describe the radiative direct neutron capture process ${\rm (A - 1) + n \to A^{*} \to A^{**} + \gamma}$
taking place via collective and non-collective states ${\rm A^{*}}$ decaying to low-lying excited states ${\rm A^{**}}$ 
of the synthesized nucleus. We shall proceed in two steps. As the first step, given in the present publication, we formulate a new method
to calculate the transition matrix elements of the gamma transitions between two excited states  ${\rm A^{*}}$
and ${\rm A^{**}}$. Calculation of the transition matrix elements between excited states are straightforward if
both states are discrete bound states and their wave functions are explicitly given on discrete basis. 
A novel feature of the formulation proposed here is that we use the linear response theory which is able to
describe continuum excited state ${\rm A^{*}}$ located above the neutron 
separation energy. This is an essential requirement in applying to the neutron-rich nuclei near the drip-line.
The second step, an application to the radiative direct neutron-capture reaction
will be given in a forthcoming paper.

In section 2, we formulate a linear response theory extended to calculate transition matrix elements between excited states.
For this purpose we define a non-local one-body operator introduced to evaluate the matrix elements. Applying the linear response formalism
to this non-local operator, we obtain a new type of strength function which describes excitation modes in the continuum
and  transition matrix elements with respect to a low-lying excited state. This extended linear response formalism
enables us to evaluate the branching ratios for gamma-decays from the continuum excited states to different 
low-lying excited states as well as the ground states. In section 3, we 
demonstrate applicability of the present approach by performing a numerical calculation for a neutron-rich nucleus
$^{140}{\rm Sn}$, and discuss the dipole, quadrupole and octupole excitations including the continuum particle-hole
modes and the giant resonances, and the transitions from/to low-lying $2^+$ and $3^{-}$ states. 
We draw conclusions in section 5. 

\section{Theory}

In the first three subsections we introduce the extended formalism of the 
continuum random-phase approximation which describes the transitions between RPA excited states.
We then provide, in the following subsections, a detailed formulation for an application to a spherical nucleus
with a $j$-shell closed configuration.

\subsection{Strength function for transitions between RPA excited states}

We shall describe excited states $\{ \ket{k} \}$
by means of the random phase approximation (RPA) to oscillations 
 around the ground state $\ket{0}$. The standard RPA formalism provides a
  scheme to calculate the transition matrix elements $\bra{k} \hat{M} \ket{0}$ 
  and the strength function 
  $S(\hat{M}; \hbar\omega) \equiv \sum_{k}  | \bra{k} \hat{M} \ket{0} |^{2} \delta(\hbar \omega - (E_{k} - E_{0}))$
  for a one-body operator $\hat{M}$, e.g.,  an electro-magnetic multipole operator\cite{Ring1980}.
  
  In the present paper, we consider another RPA excited state $\ket{i}$, for instance, the low-lying
  collective state with  a character of surface vibration, and we intend to describe the transition matrix elements
   $\bra{k} \hat{M} \ket{i}$ between the low-lying state $\ket{i}$ and
   the RPA excited states $\{ \ket{k} \}$ under consideration. 
   Since we consider neutron-rich (or proton-rich) nuclei and 
   the situation where the RPA excited states $\{ \ket{k} \}$ are populated
   via the direct neutron (proton) capture reaction, we shall treat the RPA excited states $\{ \ket{k} \}$  as those
embedded in the continuum spectrum above the neutron (proton) separation energy. It is then appropriate to
describe $\{ \ket{k} \}$ by means of the continuum RPA, i.e., the linear response theory using 
the Green's function technique\cite{Bertsch1975,Shlomo1975}.
   
We introduce a strength function for transitions between RPA excited states $\ket{i}$ and $\{ \ket{k} \}$: 
\begin{align}
	S(\hat{M}; i ; \Delta E) = \sum_{k} | \bra{k} \hat{M} \ket{i} |^{2} \delta(\Delta E - (E_{k} - E_{i})).
\label{S_M_i_org}
\end{align}
Here $\ket{i}$ is fixed  and $\ket{k}$ runs over all excited states described by the
continuum RPA.  The RPA excited states are generally described in terms of the mode creation operator 
  $\hat{O}^{\dag}$, a linear combination of the particle-hole and hole-particle excitations, which is written as
 \begin{align}
	\hat{O}^{\dag}_{i} = \sum_{ph} \left\{ X^{i}_{ph} a^{\dag}_{p} a_{h} - Y^{i}_{ph} a^{\dag}_{h} a_{p} \right\},
\end{align}
e.g. for the state $\ket{i}=\hat{O}^{\dag}_{i}\ket{0}$.
Using the mode creation operator, the strength function (\ref{S_M_i_org}) can be rewritten as
\begin{align}
	S(\hat{M}; i ; \hbar \omega)  &\equiv \sum_{k} | \bra{k} \hat{M} \ket{i} |^{2} \delta(\hbar \omega - (E_{k} - E_{0})) \notag \\
	&= \sum_{k} | \bra{k} [ \hat{M} , \hat{O}^{\dag}_{i}] \ket{0} |^{2} \delta(\hbar \omega - (E_{k} - E_{0})) = S (\hat{F}; \hbar \omega).
\label{S_M_i}
\end{align}
Note that the second expression can be regarded as a strength function for transitions from the ground state $\ket{0}$ caused by a newly defined operator
$\hat{F}\equiv [\hat{M}, O^{\dag}_{i}]$. The replacement by the commutator is valid under  
the quasi-boson approximation\cite{Ring1980} $[\hat{O}^{\dag}_{i}, \hat{O}_{k}] = \delta_{ik}$, which is
 equivalent to keeping the leading orders of the boson expansion of $\hat{M}, \hat{O}$, and $\hat{O}^{\dag}$.
For simplicity  the excitation energy $\hbar\omega=E_k - E_0$ is used in Eq.(\ref{S_M_i}) 
in place of the transition energy $\Delta E=E_k- E_i$ in Eq.(\ref{S_M_i_org}).

We remark here that the commutator $\hat{F}= [\hat{M}, O^{\dag}_{i}] $ 
is a one-body but {\it non-local} operator.  In fact, for the multipole moment 
\begin{align}
	\hat{M} = \int dx {f}(\mathbf{r}_{x}) \hat{\rho}(x), \quad f(\mathbf{r}_{x}) \equiv r^{L}_{x} Y_{L M} (\Omega_{x}),
\end{align}
the operator $\hat{F}$ is
\begin{align}
	\hat{F} = \iint dx dy F(x, y) \psi^{\dag}(x) \psi(y)
\end{align}
\begin{align}
	F(x, y) \equiv (f(\mathbf{r}_{x}) - f(\mathbf{r}_{y})) \sum_{ph} \left\{ X^{i}_{ph} \phi_{p}(x) \phi^{*}_{h}(y) - 
	Y^{i}_{ph} \phi_{h}(x) \phi^{*}_{p}(y) \right\}.
\label{F_matel}
\end{align}
Here $\psi^{\dag}(x)$, $\psi(x)$ and $\hat{\rho}(x)=\psi^{\dag}(x)\psi(x)$ are the creation, annihilation and  
density operators of nucleon with a shorthand notation of the coordinate and the spin variables $x \equiv (\mathbf{r}_{x}, \sigma_{x})$
while $\phi_{p}(x)$ and $\phi_{h}(x)$ are single-particle wave functions of the particle and hole orbits, respectively.
The isospin variable is omitted for simplicity. 
The integral $\int dx$ represents $\int dx \equiv \sum_{\sigma_{x}} \int d \mathbf{r}_{x}$. 

The expression (\ref{S_M_i}) allows us to formulate the linear response of the system against an external perturbation
provided by the non-local one-body operator $\hat{F}=[\hat{M}, O^{\dag}_{i}]$. A new feature is that we need to consider a response
of the non-local density, i.e. the density matrix
$ \rho(x,y) =\langle \hat{\rho}(x, y) \rangle$ with  $\hat{\rho}(x, y) = \psi^{\dag}(y) \psi(x)$.
The response in the frequency domain is given by 
 \begin{align}
	\delta \rho(x, y, \omega)  = \iint dx^{'}dy^{'} \, R(x, y; y^{'}, x^{'}; \omega) F(x^{'}, y^{'})
\end{align}
with a response function generalized to the density matrix, which is formally expressed as
\begin{align}
	R(x, y; y^{'}, x^{'}; \omega) \equiv \sum_{k} \left\{ \frac{\bra{0} \hat{\rho}(x, y) \ket{k} \bra{k} \hat{\rho}(y^{'}, x^{'}) \ket{0}}
			{\hbar \omega - \hbar\omega_{k}  + i \eta} - \frac{\bra{0} \hat{\rho}(y^{'}, x^{'}) \ket{k} \bra{k} \hat{\rho}(x, y) \ket{0}}{\hbar 		\omega + \hbar\omega_k + i \eta} \right\}.
\end{align}
Here $\eta$ is a positive infinitesimal constant and $\hbar\omega_k=E_k-E_0$ is the excitation energy of the RPA excited states
$\left\{  \ket{k} \right\}$.
The strength function $S(\hat{F};\hbar\omega)$ in Eq.(\ref{S_M_i}) is given by 
\begin{align}
	S(\hat{F};\hbar\omega)  =  - \frac{1}{\pi} {\rm Im} \iint dxdy \, F^{*}(x, y) \delta \rho(x, y, \omega),
\label{S_F_cRPA}
\end{align}
using the density-matrix response $\delta \rho(x, y, \omega)$. 

Note that the strength function can be expressed also as
\begin{align}
	S(\hat{F};\hbar\omega) &= - \frac{1}{\pi} {\rm Im} \int dx f^{*}(x) \left\{ \int dy \bar{\rho}^{(\mathrm{tr})*}_{i}(x, y) \delta {\rho}(x, y, \omega)
	 - \int dy \bar{\rho}^{(\mathrm{tr})*}_{i}(y, x) \delta {\rho}(y, x, \omega) \right\},
\label{S_F_cRPA_form}
\end{align}
obtained by inserting Eq.(\ref{F_matel}) into Eq.(\ref{S_F_cRPA}). Here we introduced a quantity 
 \begin{align}
	\bar{\rho}^{(\mathrm{tr})}_{i}(x, y) &\equiv \sum_{ph} \left\{ X^{i}_{ph} \phi_{p}(x) \phi^{*}_{h}(y) - Y^{i}_{ph} \phi_{h}(x) \phi^{*}_{p}(y) \right\}
\label{pseudo_density_matrix}	
\end{align}
to represent the second factor in Eq.(\ref{F_matel}) associated with the RPA state $\ket{i}$.
 We call it the {\it pseudo transition density-matrix} of $\ket{i}$ since 
 it has the same structure as the transition density-matrix 
\begin{align}
	\rho^{(\mathrm{tr})}_{i}(x, y) \equiv \bra{0} \hat{\rho}(x,y) \ket{i} = 
	\bra{0} [\hat{\rho}(x,y) , \hat{O}_{i}^{\dagger} ] \ket{0}
	= \sum_{ph} \left\{ X^{i}_{ph} \phi_{p}(x) \phi^{*}_{h}(y) + Y^{i}_{ph} \phi_{h}(x) \phi^{*}_{p}(y) \right\}
\end{align}
except the sign of the second term related to the backward amplitudes $Y^{i}_{ph}$.

\subsection{Extended linear response equation}

Since the operator  $\hat{F}$ is a one-body, though non-local, operator, it is possible to formulate the linear response 
on the basis of the time-dependent Kohn-Sham theory, or
the time-dependent Hartree-Fock theory. Separating the time-dependent
selfconsistent field $U[\rho]=U_0 + U_{ind}$ into the stationary part $U_0$ associated with the ground state 
and the induced field $U_{ind} = \left(\frac{\delta U}{\delta \rho}\right)\delta\rho$, 
the density-matrix response is given by
 \begin{align}
	\delta \rho(x, y, \omega) = \iint dx^{'} dy^{'} R_{0}(x, y; y^{'}, x^{'}; \omega) (v_{ind}(x^{'}, y^{'}, \omega)+F(x^{'}, y^{'}))
\end{align}
in terms of the unperturbed response function
\begin{align}
	R_{0}(x, y; y^{'}, x^{'}; \omega) \equiv 
	\sum_{ph} \left\{ \frac{\bra{0} \hat{\rho}(x, y) \ket{ph} \bra{ph} \hat{\rho}(y^{'}, x^{'}) \ket{0}}
		{\hbar \omega - (\epsilon_{p} - \epsilon_{h}) + i \eta} - \frac{\bra{0} \hat{\rho}(y^{'}, x^{'}) \ket{ph} \bra{ph} \hat{\rho}(x, y) \ket{0}}{\hbar
		 		\omega + (\epsilon_{p} - \epsilon_{h}) + i \eta} \right\},
\end{align}
for uncorrelated particle-hole states $\ket{ph}=a^{\dag}_pa_{h}\ket{0}$. 
The unperturbed response function is also given as 
\begin{align}
R_{0}(x, y; y^{'}, x^{'}; \omega)=
	\sum_{\epsilon_{i} < \epsilon_{f}} &\left\{ \phi^{*}_{i}(y)G_{0}(x, x^{'}, \epsilon_{i} + \hbar 
				\omega + i\eta)\phi_{i}(y^{'}) \right. \notag \\ 
				&\left. + \phi^{*}_{i}(x^{'})G_{0}(y^{'}, y, \epsilon_{i} - \hbar \omega - i\eta)\phi_{i}(x) \right\},
\label{unp_respfn_G}
\end{align}
using 
the single-particle Green's function $G_{0}(x, x^{'} ,e)=\bra{x} (e - \hat{h}_0)^{-1}\ket{x^{'}}
=\sum_i \phi^{*}_{i}(x)\phi_{i}(x^{'})(e - \epsilon_{i})^{-1}$ for the mean-field Hamiltonian $\hat{h}_0 = \hat{t} + U_0$. 
The Green's function allows one to describe the single-particle states in the continuum and hence RPA excited states embedded 
in the continuum spectrum above the particle separation energy.

In the following, we assume that the induced field is local and spin-independent.
$v_{ind}(x^{'}, y^{'}, \omega) = v_{ind}(\mathbf{r}_{x^{'}}, \omega)\delta(x^{'} - y^{'}) = \frac{\delta U}{\delta \rho}(\mathbf{r}_{x^{'}}, \omega) \delta \rho(\mathbf{r}_{x^{'}}) \delta(\mathbf{r}_{x^{'}} - \mathbf{r}_{y^{'}}) \delta_{\sigma_{x^{'}} \sigma_{y^{'}}}$.
In this case, the density-matrix response $\delta \rho(y, x, \omega)$ is given by
\begin{align}
	\delta \rho(x, y, \omega) = \int dx^{'} \, R_{0}(x, y; x^{'}, x^{'}; \omega) \frac{\delta U}{\delta \rho}(\mathbf{r}_{{x}^{'}}) \delta 	\rho (\mathbf{r}_{x^{'}}, \omega) 
	+ \iint dx^{'} dy^{'} R_{0}(x, y; y^{'}, x^{'}; \omega) F(x^{'}, y^{'}).
\label{general_linear_response_eq}
\end{align}
For the  local density response $\delta \rho(\mathbf{r}_{x}, \omega)=\sum_{\sigma_{x}} \delta \rho(x, y=x, \omega)$
appearing in the right hand side of Eq.(\ref{general_linear_response_eq}), we have an integral equation
\begin{align}
	\delta \rho(\mathbf{r}_{x}, \omega) 
	= \sum_{\sigma_{x}} \int dx^{'} R_{0}(x, x; x^{'}, x^{'}; \omega) \frac{\delta U}{\delta \rho}(\mathbf{r}_{x^{'}}) \delta 			
	\rho(\mathbf{r}_{x^{'}}, \omega) + \sum_{\sigma_{x}} \iint dx^{'}dy^{'} R_{0}(x, x; y^{'}, x^{'}; \omega) F(x^{'}, y^{'}).
\label{local_linear_response_eq}
\end{align}

We can solve numerically the integral equation (\ref{local_linear_response_eq}) by treating it as a linear equation on the mesh 
points in the coordinate space. 
Inserting the density response $\delta \rho(\mathbf{r}_{x}, \omega)$ into Eq.(\ref{general_linear_response_eq}), 
the density-matrix response $\delta \rho(x, y, \omega)$ is obtained,  
and finally we can calculate the strength function $S(\hat{F}; \hbar\omega)$
using Eq.(\ref{S_F_cRPA_form}). Note that the pseudo transition density-matrix  $\bar{\rho}^{(\mathrm{tr})}_{i}(x, y)$ of the state $\ket{i}$ can be calculated
also within the linear response formalism (the continuum RPA formalism) as we discuss in Appendix A. Consequently
all the calculations are done within the consistent framework of the continuum RPA.

\subsection{Transition densities and diagrammatic interpretation}

We first note that the transition densities for transitions between the ground state and
the RPA excited states are calculated as 
\begin{align}
\rho^{(\mathrm{tr})}_{k}(x) \equiv \bra{0} \hat{\rho}(x) \ket{k} = C \mathrm{Im} \delta\rho(x,\omega_k)
\end{align}
for the local transition density, and similarly
\begin{align}
\rho^{(\mathrm{tr})}_{k}(x,y) \equiv \bra{0}\psi^{\dagger}(y)\psi(x) \ket{k}= C \mathrm{Im} \delta\rho(x,y,\omega_k)
\end{align} 
for the transition density-matrix.
The density responses $\delta\rho(x,\omega_k)$ and $\delta\rho(x,y,\omega_k)$ are solutions of
Eqs.(\ref{general_linear_response_eq})  and (\ref{local_linear_response_eq})
at the excitation energy $E_k-E_0=\hbar\omega_k$ of the state $\ket{k}$.
Here $C$ is a
normalization constant which is determined to reproduce the transition strength evaluated 
from the strength function. 

Now we shall consider the transition density for the transition between the RPA excited states, i.e. 
the one between $\ket{i}$ and $\ket{k}$: 
\begin{align}
\rho^{(\mathrm{tr})}_{i,k}(x) \equiv \bra{i} \hat{\rho}(x) \ket{k} =  \bra{0}[\hat{O}_i, \hat{\rho}(x)] \ket{k} . 
\end{align}
Using the commutation relation
\begin{align}
[\hat{O}_i, \hat{\rho}(x)]=\int dy \left\{ \bar{\rho}^{(\mathrm{tr}) *}_{i}(x,y) \psi^{\dagger}(y)\psi(x) -  \bar{\rho}^{(\mathrm{tr}) *}_{i}(y,x) \psi^{\dagger}(x)\psi(y) \right\} , 
\end{align}
the transition density
is given as
\begin{align}
\rho^{(\mathrm{tr})}_{i,k}(x) = \int dy  \left\{ \bar{\rho}^{(\mathrm{tr}) *}_{i}(x,y) \rho^{(\mathrm{tr})}_{k}(x,y) - \bar{\rho}^{(\mathrm{tr})  *}_{i}(y,x) \rho^{(\mathrm{tr})}_{k}(y,x) \right\}
\end{align}
expressed as a convolution of the  transition density-matrix 
$\rho^{(\mathrm{tr})}_{k}(x,y)$  
for the state $\ket{k}$ and the
pseudo transition density-matrix $\bar{\rho}^{(\mathrm{tr})}_{i}(x,y)$  for the state $\ket{i}$. 

\begin{figure}
	\centering
	\includegraphics[width=0.5\columnwidth]{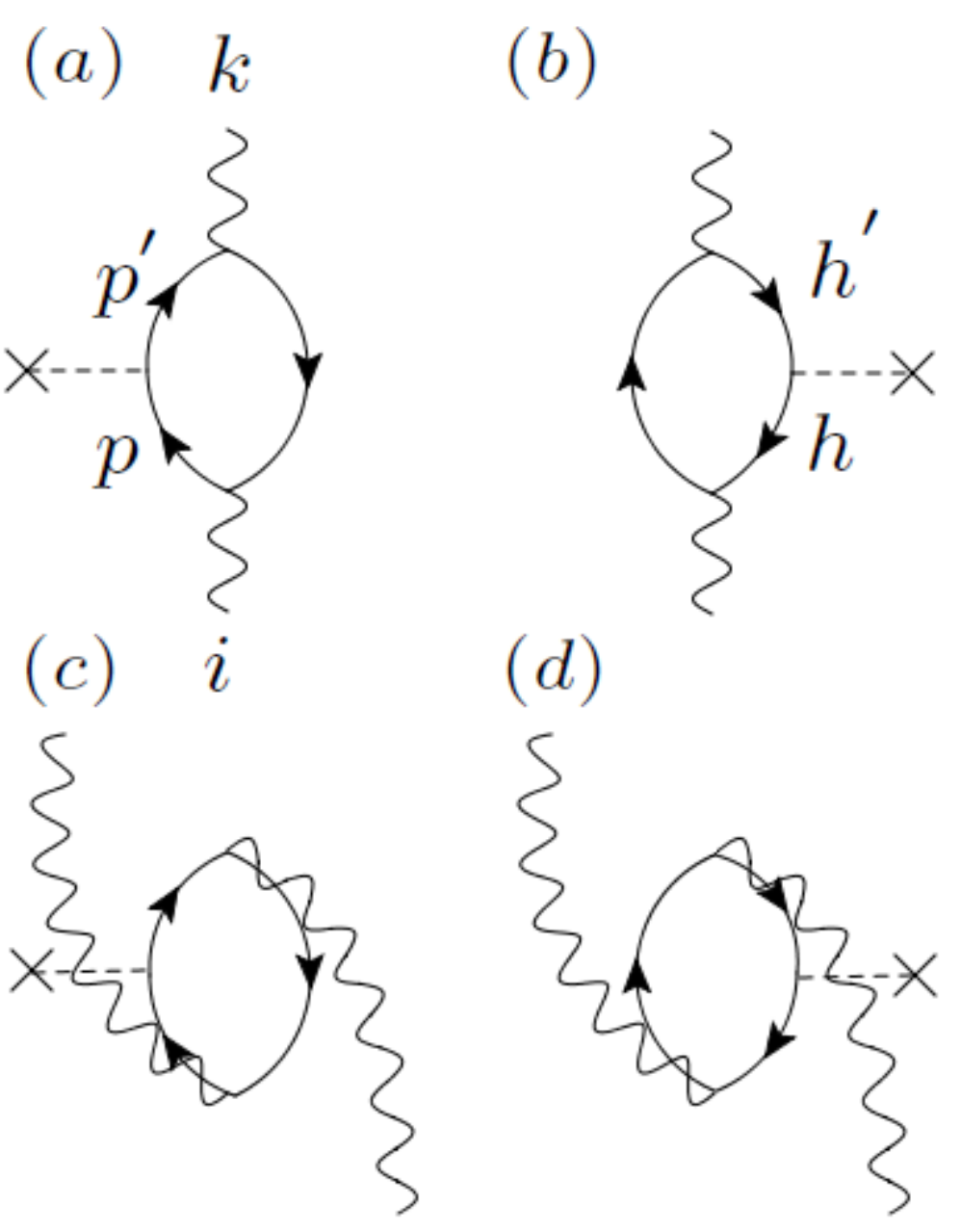}
	\caption{The diagram representation of the transition density or the matrix element of a one-body operator
	between RPA excited states $\ket{i}$ and $\ket{k}$. See Eq.(\ref{trans_xy}) and the text.}
	\label{diagram_fb}
\end{figure}

The transition density is expressed also
in terms of the forward and backward amplitudes of the mode creation operators:
\begin{align}
\rho^{(\mathrm{tr})}_{i,k}(x)  = \sum_{pp'h}X_{ph}^{i *}X_{p'h}^{k}\phi_{p}^{*}(x)\phi_{p'}(x) 
- \sum_{phh'}X_{ph}^{i *}X_{ph'}^{k}\phi_{h}(x)\phi_{h'}^{*}(x) \notag \\
+\sum_{pp'h}Y_{ph}^{i *}Y_{p'h}^{k}\phi_{p}^{*}(x)\phi_{p'}(x) 
- \sum_{phh'}Y_{ph}^{i *}Y_{ph'}^{k}\phi_{h}(x)\phi_{h'}^{*}(x) .
\label{trans_xy}
\end{align} 
It is possible to interpret each term using a diagrammatic representation  as shown in
Fig.\ref{diagram_fb}.  Figure \ref{diagram_fb} (a) and (b), corresponding to the first and second terms of Eq.(\ref{trans_xy}), represent
actions of the operator on particle and hole components of the RPA states, respectively
whereas Fig. \ref{diagram_fb} (c) and (d) are counterparts, the third and fourth terms, associated with the backward amplitudes of the RPA
states.

The transition matrix elements $\bra{i} \hat{M} \ket{k}=\int dx f(x) \rho^{(\mathrm{tr})}_{i,k}(x) $ 
between the RPA excited states for a one-body
operator $\hat{M}$  is also
represented in terms of the same diagrams. 

\subsection{Spherical system}

In this section, we give explicit formulae which can be used in actual numerical calculation. Here
the spherical symmetry of the ground state and the associated mean-field is assumed.

Suppose that the ground state $\ket{0}$, the RPA excited states $\ket{k}$, $\ket{i}$ and
the transition operator $\hat{M}$ have  the angular momentum quantum numbers
  $0^{+}$,  $L M$, $L_{i}M_{i}$ and $\lambda \mu$, respectively. 
The operator $\hat{F}= [\hat{M}, \hat{O}^{\dag}_{i}]$ is given the explicit rank  $LM$:
\begin{align}
	\hat{F}_{L M} \equiv \sum_{\mu M_{i}} \langle \lambda \mu L_{i} M_{i} | L M \rangle [\hat{M}_{\lambda \mu}, \hat{O}^{\dag}_{iL_{i}M_{i}}] .
\end{align} 
Using this operator we evaluate the strength function for the transitions from the ground state $ \ket{0^{+}_{\mathrm{g}}}$ to excited 
RPA states $\ket{kLM}$
\begin{align}
S(\hat{F}_{L}; \mathrm{g}, L ; \hbar \omega) 
&\equiv \sum_{k M} |\bra{k L M} \hat{F}_{L M} \ket{0^{+}_{\mathrm{g}}}|^{2} \delta(\hbar \omega - (E_{k} - E_{0})) \notag \\
&= \sum_{k} |\bra{k L} |\hat{F}_{L} |\ket{0^{+}_{\mathrm{g}}}|^{2} \delta(\hbar \omega - (E_{k} - E_{0})).
\label{S_FL}
\end{align}
It is identical to the strength function  
\begin{align}
S(\hat{M}_{\lambda}; iL_{i}, L ; \hbar\omega) \equiv \sum_{k} |\bra{k L} |\hat{M}_{\lambda}| \ket{i L_{i}}|^{2} \delta(\hbar \omega - (E_{k} - E_{0}))=
S(\hat{F}_{L}; \mathrm{g}, L ; \hbar \omega) 
\label{S_MLiL}
\end{align}
which describes reduced matrix elements for transitions from the RPA excited state $\ket{i L_{i} M_{i}}$ 
to a set of RPA excited states $\left\{ \ket{k L M} \right\} $. Note that the angular quantum numbers $L_i$ and $L$ of the excited states are
explicitly indicated to label the strength functions (\ref{S_FL}) and (\ref{S_MLiL}).

The density response and the density-matrix response caused by $\hat{F}_{L M}$ also have quantum numbers $L M$. These functions and 
the matrix element of $\hat{F}_{L M}$ are expanded by the spherical harmonics and the spin spherical harmonics as
\begin{align}
	\delta \rho(\mathbf{r}_{x}, \omega) = Y_{L M}(\hat{\mathbf{r}}_{x}) \frac{1}{r^{2}_{x}} \delta \rho_{L}(r_{x}, \omega),
\end{align}
\begin{align}
	\delta \rho(x, y, \omega) = \sum_{ljm, l^{'}j^{'}m^{'}} Y_{l^{'}j^{'}m^{'}}(\hat{x}) \frac{1}{\sqrt{2j^{'} + 1}} \langle j m L M| j^{'} m^{'} \rangle \frac{\delta \rho_{L, l^{'}j^{'}, lj}(r_{x}, r_{y}, \omega)}{r_{x} r_{y}}Y^{*}_{ljm}(\hat{y}),
\end{align}
\begin{align}
	F_{L M}(x, y) = \sum_{ljm, l^{'}j^{'}m^{'}} Y_{l^{'}j^{'}m^{'}}(\hat{x}) \frac{1}{\sqrt{2j^{'} + 1}} \langle j m L M| j^{'} m^{'} \rangle \frac{F_{L, l^{'}j^{'}, lj}(r_{x}, r_{y})}{r_{x} r_{y}} Y^{*}_{ljm}(\hat{y}).
\end{align}
The extended linear response equations  for the radial functions of the density responses are given as
\begin{align}
	\delta \rho_{L, l^{'}j^{'}, lj}(r_{x}, r_{y}, \omega) = \bra{l^{'}j^{'}} |Y_{L}| \ket{lj} &\int dr_{x^{'}} R_{0, l^{'}j^{'}, lj}(r_{x}, r_{y};r_{x^{'}}, r_{x^{'}}; \omega) \frac{\delta U}{\delta \rho}(r_{x^{'}}) \frac{1}{r^{2}_{x^{'}}} \delta \rho_{L}(r_{x^{'}}, \omega) 		\notag \\
+ &\iint dr_{x^{'}} dr_{y^{'}} R_{0, l^{'}j^{'}, lj}(r_{x}, r_{y}; r_{y^{'}}, r_{x^{'}}; \omega) F_{L, l^{'}j^{'}, lj}(r_{x^{'}}, r_{y^{'}}),
\label{radial_delta_rhomat}
\end{align}
\begin{align}
	\delta \rho_{L}(r_{x}, \omega) = \sum_{lj, l^{'}j^{'}} \biggl\{ \frac{|\bra{l^{'}j^{'}} |Y_{L}| \ket{lj}|^{2}}{2L + 1} &\int dr_{x^{'}} R_{0, l^{'}j^{'}, lj}(r_{x}, r_{x}; r_{x^{'}}, r_{x^{'}}; \omega) \frac{\delta U}{\delta \rho}(r_{x^{'}}) \frac{1}{r^{2}_{x^{'}}} \delta \rho_{L}(r_{x^{'}}, \omega) \notag \\
+ \frac{\bra{l^{'}j^{'}} |Y_{L}| \ket{lj}^{*}}{2L + 1} &\iint dr_{x^{'}} dr_{y^{'}} R_{0, l^{'}j^{'}, lj}(r_{x}, r_{x}; r_{y^{'}}, r_{x^{'}}; \omega) F_{L, l^{'}j^{'}, lj}(r_{x^{'}}, r_{y^{'}}) \biggr\}.
\label{radial_delta_rho}
\end{align}

The explicit form of radial unperturbed response function $R_{0, l^{'}j^{'}, lj}$ is given in Appendix A
and can be calculated using the exact single-particle Green's function. 
Note also that there holds a relation
\begin{align}
	\delta \rho_{L}(r_{x}, \omega) = \frac{1}{2L + 1} \sum_{lj, l^{'}j^{'}} \bra{l^{'}j^{'}} |Y_{L}| \ket{lj}^{*} \delta \rho_{L, l^{'}j^{'}, lj}(r_{x}, r_{x}, \omega).
\end{align}

The strength function $S(\hat{F}_{L}; \mathrm{g}, L; \hbar\omega )$ is given by
\begin{align}
	S(\hat{F}_{L} ; \mathrm{g}, L;  \hbar \omega)  = & - \frac{1}{\pi} {\rm Im} \iint dr_{x} dr_{y} \sum_{\substack{lj \\ l^{'}j^{'}}} F^{*}_{L,l^{'}j^{'}, lj}(r_{x}, r_{y}) \delta \rho_{L, l^{'}j^{'}, lj}(r_{x}, r_{y}, \omega) \notag \\
		 = &- \frac{1}{\pi} {\rm Im} \int dr_{x} f^{*}_{\lambda} (r_{x}) \notag \\
	&\times \sum_{lj, l^{'}j^{'}, l_{2}j_{2}} \Bigl[ \sqrt{2L + 1} (-)^{L + 1} (-)^{j - j^{'}}  \begin{Bmatrix} \lambda & L_{i} & L \\ j & j^{'} & j_{2} \end{Bmatrix} \bra{l^{'}j^{'}} |Y_{\lambda}| \ket{l_{2}j_{2}}^{*} \notag \\
	&\qquad \qquad \qquad \times \int dr_{y} \, \bar{\rho}^{(\mathrm{tr})*}_{i L_{i}, l_{2}j_{2}, lj}(r_{x}, r_{y}) \delta \rho_{L, l^{'}j^{'}, lj}(r_{x}, r_{y}, \omega) \notag \\
	&\qquad \quad \, \, + \sqrt{2L + 1} (-)^{L_{i} - \lambda} (-)^{j^{'} - j}  \begin{Bmatrix} \lambda & L_{i} & L \\ j^{'} & j & j_{2} \end{Bmatrix} \bra{l_{2}j_{2}} |Y_{\lambda}| \ket{lj}^{*} \notag \\
	&\qquad \qquad \qquad \times \int dr_{y} \, \bar{\rho}^{(\mathrm{tr})*}_{ i L_{i}, l^{'}j^{'}, l_{2}j_{2}}(r_{y}, r_{x}) \delta \rho_{L, l^{'}j^{'}, lj}(r_{y}, r_{x}, \omega) \Bigr].
\end{align}
Here is used the expression for the matrix element of the operator $\hat{F}_{LM}$ 
\begin{align}
	F_{L, l^{'}j^{'}, lj}(r_{x}, r_{y}) = \sum_{l_{2}j_{2}} \sqrt{2L + 1} \Bigl[ (-)^{L + 1} (-)^{j - j^{'}} &\begin{Bmatrix} \lambda & L_{i} & L \\ j & j^{'} & j_{2} \end{Bmatrix} \bra{l^{'}j^{'}} |Y_{\lambda}| \ket{l_{2}j_{2}} f_{\lambda}(r_{x}) \bar{\rho}^{(\mathrm{tr})}_{ iL_{i}, l_{2} j_{2}, l j}(r_{x}, r_{y}) \notag \\
+ (-)^{L_{i} - \lambda} (-)^{j^{'} - j} &\begin{Bmatrix} \lambda & L_{i} & L \\ j^{'} & j & j_{2} \end{Bmatrix} \bra{l_{2} j_{2}} |Y_{\lambda}| \ket{lj} f_{\lambda}(r_{y}) \bar{\rho}^{(\mathrm{tr})}_{ i L_{i}, l^{'}j^{'}, l_{2}j_{2}}(r_{x}, r_{y})  \Bigr],
\label{radical_F_cRPA_form}
\end{align}
with the pseudo radial transition density-matrix $\bar{\rho}^{(\mathrm{tr})}_{ iL_{i}, l^{'}j^{'}, lj}$ for the low-lying discrete state $\ket{iL_iM_i}$.

\subsection{Photoemission decays to low-lying states}

One can apply the above formulation to describe
photoemission transitions from excited states in the continuum to a low-lying excited state.
We consider a transition  of multipole $\lambda$ from the excited state $\ket{kLM}$ in the continuum at energy $E_k$ to the
low-lying bound excited state $\ket{iL_i M_i}$ at $E_i$.   The transition probability\cite{Ring1980}, proportional to the
reduced matrix element $B(M_{\lambda}, kL \to iL_{i}) = \frac{1}{2L + 1} |\bra{i L_{i}} |\hat{M}_{\lambda} |\ket{k L}|^{2} \Delta E$,
is given by
\begin{align}
	T_{kL \to iL_i} &= \frac{8 \pi (\lambda + 1)}{\hbar \lambda ((2 \lambda + 1)!!)^{2}} \left( \frac{E_{\gamma}}{\hbar c} \right)^{2 \lambda + 1} \frac{1}{2L + 1}  S(\hat{M}_{\lambda}; iL_{i}, L; E_{k} - E_{0}) \Delta E
\end{align}
with $E_\gamma = E_{k} - E_{i}$ using the strength function $S(\hat{F}_{L}; \mathrm{g}, L; E_k-E_0)=S(\hat{M}_{\lambda}; iL_i, L ; E_k-E_0)$.  
The energy interval $\Delta E$ is chosen arbitrarily small for the continuum $\ket{k}$  whereas
in the case of discrete $\ket{k}$
it should be treated as an integral $\int^{E_{k} + \Delta E/2}_{E_{k} - \Delta E/2} S(\hat{F}_{L}; \mathrm{g},L; E - E_{0}) dE$ 
to cover the associated peak structure of the strength function.

\section{Numerical example}

\subsection{Setting}

We shall describe electromagnetic transitions in neutron-rich nucleus
$^{140}{\rm Sn}$ in order to demonstrate the present theory.  
The neutron separation energy in this nucleus is predicted as small as $S_n \sim 3 $MeV by the
Hartree-Fock calculations~\cite{Massexpl}, and it may be one of the isotopes which play roles in the r-process neutron-capture. We notice also
that the pair correlation of neutrons 
is expected to be weak due to a single-$j$ closed configuration.

We focus on excited states with spin-parity $1^{-}$, $2^{+}$ and $3^{-}$ where characteristic excited states are expected
to emerge both in low-lying and high-lying regions. Examples are the soft dipole mode and the giant dipole resonance for
$1^{-}$,  the low-lying  quadrupole
state and the isoscalar/isovector giant quadrupole resonances for $2^{+}$, and the low-lying octupole collective states for $3^{-}$
as well as continuum
particle-hole excitations  above the neutron separation energy. We shall discuss electric multipole transitions (
E1, E2 and E3)  which occur among these states and the ground state. 

The numerical calculations is performed with the following setting.
We use a Woods-Saxon potential in place of the Hartree-Fock mean-field $U_0$ and a Skyrme-type contact interaction
as the residual two-body force $v_{ph}=\delta U/\delta \rho$,  given by
\begin{align}
v_{ph}(\mathbf{r}, \mathbf{r}^{'}) = \left\{t_{0}(1 + x_{0} P_{\sigma}) + \frac{t_{3}}{12}(1 + x_{3} P_{\sigma}) \rho(r)\right\} \delta(\mathbf{r} - \mathbf{r}^{'})
\end{align}
where we adopt the same parameter as Ref.\cite{Shlomo1975}: $t_{0} = f \times (-1100) \, {\rm fm^{3} \, MeV}$,$t_{3} = f \times 16000 \, {\rm fm^{6} \, MeV}$, $x_{0} = 0.5$, $x_{3} = 1$, $P_{\sigma}$ is the spin-exchange operator. 
The Woods-Saxon parameter is that of Ref.\cite{Shlomo1975}, and the Coulomb potential for
a uniform charge sphere is included for protons.
Since 
the Woods-Saxon potential is not the self-consistent potential derived from the interaction,
we  impose an approximate self-consistency condition on this residual interaction by multiplying a renormalization factor $f=0.749$ 
to $v_{ph}$ so that the spurious mode of the center of mass motion,  appearing as a RPA eigen mode with multipole $1^{-}$,
has  zero excitation energy. 

We obtain single-particle wave function and the single-particle Green's function $G_{0}$ by solving the radial 
Schro\"{o}dinger equation with the Runge-Kutta method  up to
a maximal radius $R_{\rm max} = 20 \, {\rm fm}$ (with interval $\Delta r = 0.2 \, {\rm fm}$). At $r=R_{\rm max}$
the single-particle wave function is connected to the asymptotic wave, i.e, the
Hankel function with an appropriate (complex) wave number. All the single-particle partial waves are
included, i.e.,  up to the maximum orbital angular momentum 
 $l_{\rm \max} = l_{h, \max} + {\rm max} \{ L_{i}, L \}  + 1$ where $l_{h, \max}$ is the largest among the hole orbits.
 The small imaginary constant $\eta$ in the response equation is set to $\eta = 0.1 \, {\rm MeV}$ in most cases
 although it is chosen much smaller in specific cases.
 
\begin{figure}
	\centering
	\includegraphics[width=0.6\columnwidth]{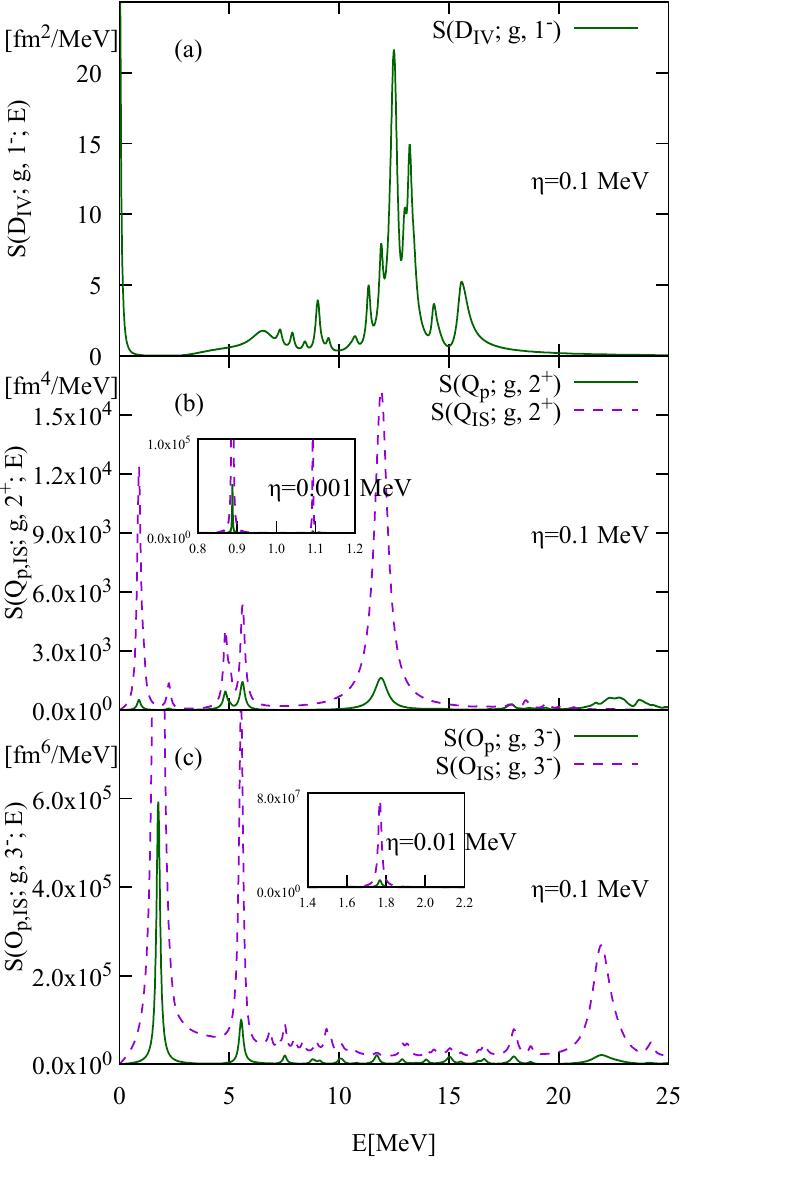}
	\caption{(a)The E1 strength function $S(D_{\mathrm{IV}}; \mathrm{g}, 1^{-}; E)$ for excited $1^{-}$ states
	in $^{140}{\rm Sn}$, calculated with $\eta=0.1 {\rm \, MeV}$. The horizontal axis is the excitation energy of the $1^{-}$ states. 
	(b)The  E2 and isoscalar quadrupole strength functions $S(Q_{\mathrm{p}}; \mathrm{g},2^{+}; E)$ and $S(Q_{\mathrm{IS}}; \mathrm{g},2^{+}; E)$ 
	for excited $2^{+}$ states, plotted with green solid and magenta dashed curves, respectively.
	The inset shows the result with $\eta=0.001 {\rm \, MeV}$, in which the $2_1^{+}$ and
	$2_2^{+}$ states are separately seen at excitation energy $E_{2^{+}_{1}} = 0.888 {\rm \, MeV}$
	and  $E_{2^{+}_{2}} = 1.093 {\rm \, MeV}$.
	(c) The E3 and isoscalar octupole strength functions
	$S(O_{\mathrm{p}}; \mathrm{g},3^{-}; E)$ and $S(O_{\mathrm{IS}}; \mathrm{g},3^{-}; E)$  for excited $3^{-}$ states. The lowest energy peak is the $3_1^{-}$
	state with $E_{3^{-}_{1}} = 1.768 {\rm \, MeV}$.
	}
	\label{sf_gs}
\end{figure}

 Figure \ref{sf_gs} shows the strength functions for the transitions from the ground state
 to the excited states with spin-parity $L^{\pi}=1^{-}, 2^{+}$ and $3^{-}$: 
 (a) the E1 strength function $S(D_{\mathrm{IV}}; \mathrm{g}, 1^{-}; E)$ for $1^{-}$ states, excited by 
 $\hat{D}_{\mathrm{IV}}= \frac{N}{A}\sum_{i,\mathrm{proton}} (rY_{1\mu})_i-\frac{Z}{A}\sum_{i,\mathrm{neutron}} (rY_{1\mu})_i$, 
 (b) the E2 and isoscalar quadrupole strength functions
 $S(Q_{\mathrm{p}}; \mathrm{g}, 2^{+}; E)$ and $S(Q_{\mathrm{IS}}; \mathrm{g}, 2^{+}; E)$ for $2^{+}$ states with
 $\hat{Q}_{\mathrm{p}}= \sum_{i,\mathrm{proton}}(r^2Y_{2\mu})_i$ and $\hat{Q}_{\mathrm{IS}}= \hat{Q}_{\mathrm{p}} + \hat{Q}_{\mathrm{n}}$,
 (c) the E3 and isoscalar octupole strength functions
 for $3^{-}$ states with
 $\hat{O}_{\mathrm{p}}=\sum_{i,\mathrm{proton}}(r^3Y_{3\mu})_i$ and  $\hat{O}_{\mathrm{IS}}= \hat{O}_{\mathrm{p}} + \hat{O}_{\mathrm{n}}$.

 Table \ref{spe_140Sn_2} shows the single-particle orbits. The energy of the neutron 
 $2f_{7/2}$ orbit (the Fermi energy ) is $e_{2f_{7/2}}= -2.59 $ MeV. As seen in Fig.\ref{sf_gs}(a), 
 there exist low-energy dipole strength which
 is discributed continuously above the neutron threshold $S_{n} = 2.59 {\rm \, MeV}$. 
 This continuum strength is
 brought mainly by the neutron particle-hole excitation 
 from  the last occupied
 $2f_{7/2}$ orbit to the continuum $d_{5/2}$ orbit. 
 (We denote this configuration as "$\nu [(\mathrm{cont.}d_{5/2})(2f_{7/2})^{-1}]_{1^{-}}$" in the following.)
 The large strength distributed around $E \approx 11-15$ MeV corresponds to the giant dipole resonance (GDR).
 From the strength functions of $2^+$ states, we focus on the 
 lowest two discrete states at $E_{2^{+}_{1}} = 0.888 {\rm \, MeV}$,  $E_{2^{+}_{2}} = 1.093 {\rm \, MeV}$
lying below the neutron threshold. The strength distributions around $E \approx 12-13$ MeV and $E \approx 21-25$ MeV are the
 isoscalar giant quadrupole resonance (ISGQR) and the isovector one (IVGQR). Two peaks around $E \approx 5$ and $\approx 6$ MeV consists
 mainly of proton particle-hole excitations $\pi [(1g_{7/2})(1g_{9/2})^{-1}]_{2^{+}}$ and 
 $\pi[(2d_{5/2})(1g_{9/2})^{-1}]_{2^{+}}$ while the enhanced isoscalar strengths
 of these peaks indicate contributions of neutron particle-hole components. For $3^{-}$ states, we notice a low-lying discrete state
 at $E_{3^{-}_{1}} = 1.768 {\rm \, MeV}$, which has a character of the octupole surface vibration.

 In the following discussion we pick up the three discrete states, the first and second $2^{+}$ states and
 the first $3^{-}$ states as the low-lying excited state $\ket{i L_{i}}$. We shall describe  
 the matrix element $\bra{i L_{i}}|\hat{M}_\lambda|\ket{k L}$ of
 multipole transitions between these low-lying states and the 
 RPA excited states $\ket{k L}$ lying above the neutron threshold in the $1^{-}$, $2^{+}$ and $3^{-}$ sectors.
We describe the E1, E2 and E3 transitions using
 the operators $\hat{M}_{\lambda}=\hat{D}_{\mathrm{IV}}$, $\hat{Q}_{\mathrm{p}}$  and $\hat{O}_{\mathrm{p}}$ with the bare charge of nucleons.
  
The present theory takees into account  the collectivity/correlation
in both the initial and final states on top of the continuum effects.
We shall demonstrate this feature by comparing two calculations in which the correlation/collectivity of the excited states
either included or neglected. We drop off the induced field $v_{ind}=(\delta U/\delta \rho) \delta\rho$ in the linear
response equations when we neglect the correlation. In this case the excited states become unperturbed
particle-hole excitations.

\begin{table}
	\centering
	\caption{Single-particle energies of the adopted Woods-Saxon potential for  $^{140}{\rm Sn}$. 
	Several orbits around the Fermi energy (indicated by  lines)
	are listed. }
	\label{spe_140Sn_2}
\begin{tabular}[t]{ c c p{5mm} c c } \hline
neutron & $\epsilon \, [{\rm MeV}]$ & & proton &  $\epsilon \, [{\rm MeV}]$ \\ \hline
			$2f_{5/2}$ &   -0.31 && $1h_{11/2}$ &  -11.40\\		
			$3p_{1/2}$ &   -0.81 && $2d_{3/2}$ &  -11.61 \\ 
			$3p_{3/2}$ &   -1.46 && $2d_{5/2}$ &  -14.06\\ 
			$1h_{9/2}$ &   -1.53  && $1g_{7/2}$ &  -15.08\\ \cline{1-2}\cline{4-5}
			$2f_{7/2}$ &   -2.59 && $1g_{9/2}$ &  -19.97\\  
			$1h_{11/2}$ &   -6.64 && $2p_{1/2}$ &  -21.75\\ 
			$3s_{1/2}$ &   -8.65 && $2p_{3/2}$ &  -23.02\\ 
			$2d_{3/2}$ &   -8.65  && $1f_{5/2}$ &  -24.81\\ 
			$2d_{5/2}$ &  -10.40 &&& \\ 
			$1g_{7/2}$ &  -10.96 &&& \\ 
			$1g_{9/2}$ &  -14.64 &&& \\ \hline
\end{tabular}
\end{table}

\subsection{$1^{-}$ states: soft dipole excitation and GDRs}

Let us first consider excited dipole states  and discuss E1 and E2 transitions from the low-lying
  $2^{+}_{1,2}$ and $3^{-}_{1}$ states.

\subsubsection{E1 transition $2^{+}_{1,2} \to 1^{-}$}

\begin{figure}
	\centering
	\includegraphics[width=0.6\columnwidth]{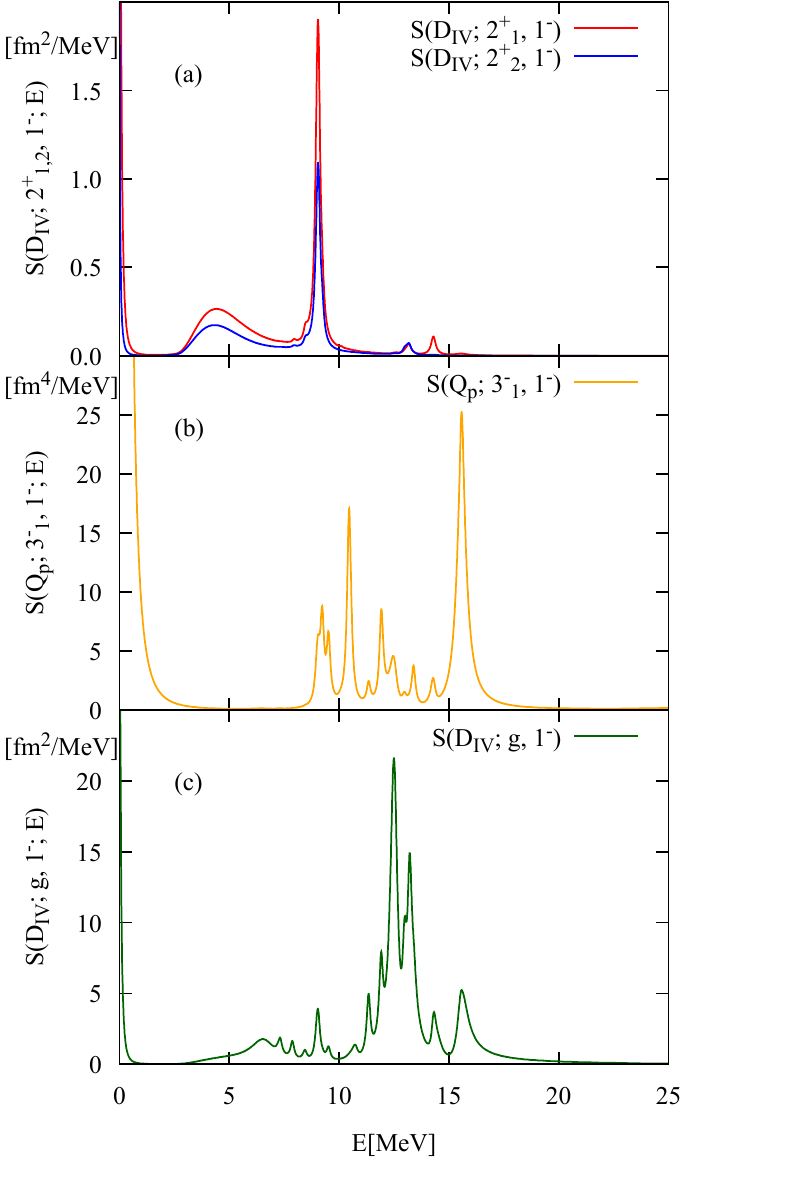}
	\caption{(a)The E1 strength functions $S(D_{\mathrm{IV}}; 2^{+}_{1,2}, 1^{-}; E)$ for transitions from $2^{+}_{1,2}$ to  $1^{-}$ states.
	(b) The E2 strength function $S(Q_{\mathrm{p}}; 3^{-}_{1}, 1^{-}; E)$ for transitions from $3^{-}_{1}$ to  $1^{-}$ states.
	(c) The E1 strength function $S(D_{\mathrm{IV}}; \mathrm{g},1^{-}; E)$ for 
	 transitions from the ground state to $1^{-}$ states. The horizontal axis is the excitation energy of the $1^{-}$ states.}
	\label{sf_multi_xt1}
\end{figure}

Figure \ref{sf_multi_xt1} (a) shows the strength function $S(D_{\mathrm{IV}}; 2^{+}_{1,2}, 1^{-};  E) $ for the E1 transitions
from the low-lying $2^{+}_{1,2}$ states to excited $1^{-}$ states. The strength function  
$S(D_{\mathrm{IV}}; \mathrm{g}, 1^{-};  E)$ for the E1
transitions from the ground state is also plotted in the lower panel for comparison.
It is seen that the strength distribution for the transitions from  $2^{+}_1$ and $2^{+}_2$
is very different from that from the ground state. We note here that 
the strength function $S(D_{\mathrm{IV}}; 2^{+}_{1,2}, 1^{-};  E) $ has little
strength in the GDR region ($E \approx 12-17$ MeV) while there exists a rather sharp peak around $E \approx 9$ MeV.
We note also continuous distribution of the strength for the $1^{-}$ states
in the region of the soft dipole excitation $S_{1n}= 2.59 < E  \lesssim 7$ MeV. However the
shape of this continuum strength is different from 
that in 
 the E1 strength function $S(D_{\mathrm{IV}}; \mathrm{g},1^{-}; E)$ for the transition from the ground state.
 
\begin{figure}
	\centering
	\includegraphics[width=0.6\columnwidth]{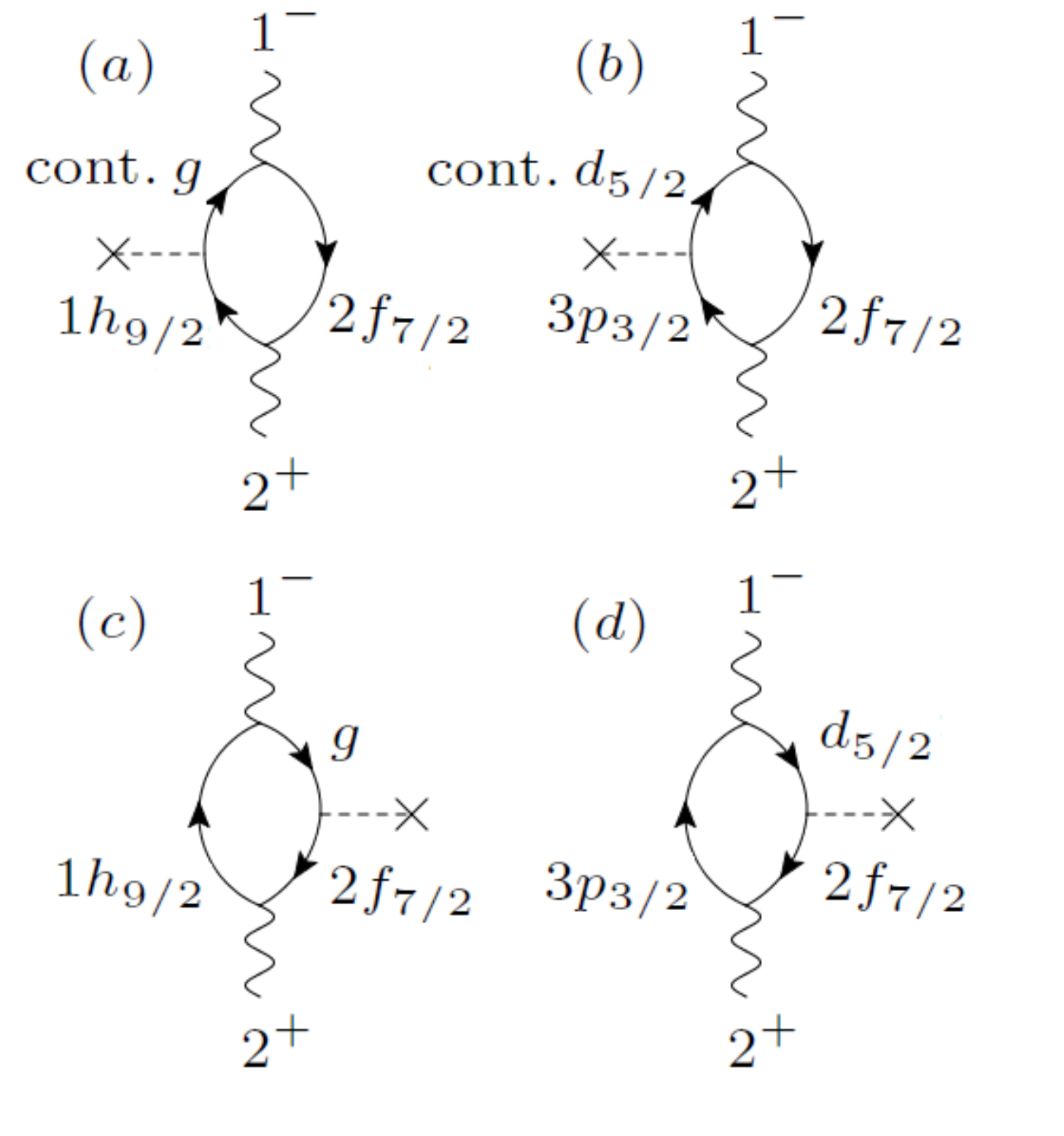}
	\caption{The diagrams representing dominant components of transition between the exited $1^{-}$  states and the low-lying $2^{+}_{1,2}$  in $^{140}{\rm Sn}$.} 
	\label{diagram_2t1}
\end{figure}

\begin{figure}
	\centering
	\includegraphics[width=0.6\columnwidth]{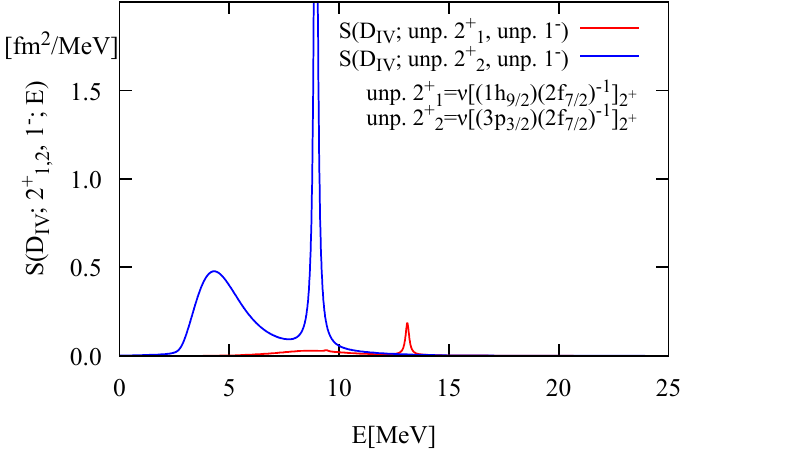}
	\caption{The E1 strength functions 
	$S(D_{\mathrm{IV};} \nu [(1h_{9/2})(2f_{7/2})^{-1}]_{2^+}, 1^{-}; E)$ and $S(D_{\mathrm{IV}}; \nu [(3p_{3/2})(2f_{7/2})^{-1}]_{2^+}, 1^{-}; E)$
	for transitions from the neutron 1p-1h states
	$\nu [(1h_{9/2})(2f_{7/2})^{-1}]_{2^+}$ and 
	$\nu [(3p_{3/2})(2f_{7/2})^{-1}]_{2^+}$ to unperturbed $1^{-}$ states.
	The horizontal axis is the excitation energy of the $1^{-}$ states. The strength function 
	$S(D_{\mathrm{IV}}; \nu [(3p_{3/2})(2f_{7/2})^{-1}]_{2^+}, 1^{-}; E)$
	 has a peak at $E=8.94$ MeV with the maximum value 3.67 $\, {\rm fm^{2}/MeV}$.
	 }
	\label{sf_multi_xt1_unp}
\end{figure}

The above behaviors can be explained with the following picture. We first note that
the correlation in the low-lying $2^{+}_{1,2}$ states is rather simple; their main structures are 
mixtures of lowest-energy neutron particle-hole excitations
  $\nu [(1h_{9/2})(2f_{7/2})^{-1}]_{2^+}$ 
and $\nu [(3p_{3/2})(2f_{7/2})^{-1}]_{2^+}$ (with excitation energies $E = 1.06 {\rm \, MeV}$ and $E= 1.13 {\rm \, MeV}$, respectively)
as indicated by the forward amplitudes shown in Table \ref{tableXamp2}.
(Note that other particle-hole configurations have small amplitudes
$| X_{ph} | < 0.1$. )
Given this feature, main components which contribute to the E1 transitions between
$2^{+}_{1,2}$ and  $1^{-}$ particle-hole excitatioins are rather limited, as is listed in Fig.\ref{diagram_2t1}. 
Figure \ref{sf_multi_xt1_unp} shows  unpertrubed
 E1 transitions associated with these components, i.e. transitions from the 
neutron particle-hole
states  $\nu[(1h_{9/2})(2f_{7/2})^{-1}]_{2^{+}}$ 
and $\nu[(3p_{3/2})(2f_{7/2})^{-1}]_{2^{+}}$ to uncorrelated $1^{-}$ particle-hole states. 
From comparison of the strength functions from $2^{+}_{1,2}$ (Fig. \ref{sf_multi_xt1} (a))
and the unperturbed strength from $\nu [(3p_{3/2})(2f_{7/2})^{-1}]_{2^+}$ (Fig. \ref{sf_multi_xt1_unp}), 
 we find that
the continuum strength in the soft dipole region $S_{1n}=2.59 < E \lesssim 7$ MeV 
originates from the component $\nu [(3p_{3/2})(2f_{7/2})^{-1}]_{2^+} \to \nu [(\mathrm{cont.}d_{5/2})(2f_{7/2})^{-1}]_{1^-}$
in which the E1 operator causes a single-particle transition of a neutron in the $3p_{3/2}$
 orbit to the continuum $d_{5/2}$ orbit. 
Absolute strengths are well explained by the mixing amplitude of  $\nu[(3p_{3/2})(2f_{7/2})^{-1}]_{2^{+}}$,
and it reflects the uncorrelated particle-hole nature of the soft dipole states.
The narrow peak around $E\approx 9$ MeV is due to 
$\nu [(3p_{3/2})(2f_{7/2})^{-1}]_{2^+} \to \nu [(3p_{3/2})(2d_{5/2})^{-1}]_{1^-}$ with
E1 transition of a neutron hole  
$(2f_{7/2})^{-1} \to (2d_{5/2})^{-1}$.
 In this case, however, 
 the strengths of this peak  deviates from simple estimation based on the
the mixing amplitudes. This is probably because the
$1^{-}$ states in this energy region is not simple particle-hole excitations.
Small peaks around $E \approx 13$ MeV  can be 
attributed to a contribution of $\nu [(1h_{9/2})(2f_{7/2})^{-1}]_{2^{+}} \to \nu[(1h_{9/2})(1g_{9/2})^{-1}]_{1^{-}}$  (Fig.\ref{diagram_2t1}(c)).
The lack of the strength  in the GDR region and higher  is a consequence of the small number of
 dominant particle-hole configurations in the low-lying $2^{+}_{1,2}$ states.
 
 \begin{table}
	\centering
	\caption{The RPA forward amplitudes $X_{ph}$ of the  $2^{+}_{1}$ and $2^{+}_{1}$ states. Particle-hole configurations with large amplitude
        $|X_{ph}| > 0.1$ are listed. 
        The RPA backward and forward amplitudes $X_{ph}$ and $Y_{ph}$ are calculated
        using a method of Ref.\cite{Shimoyama2013}}
    	\begin{tabular}{ c c c } \hline
		neutron config. \hspace{5mm}& \hspace{5mm}$X_{ph}^{2^{+}_{1}}$\hspace{5mm} & \hspace{5mm}$X_{ph}^{2^{+}_{2}}$ \hspace{5mm}  \\ \hline 
		$(1h_{9/2})(2f_{7/2})^{-1}$ & -0.601 & 0.791 \\ 
		$(3p_{3/2})(2f_{7/2})^{-1}$ & 0.789 & 0.600  \\ \hline
   	\end{tabular}
	\label{tableXamp2}
\end{table}

 \subsubsection{E2 transition $3^{-}_{1} \to 1^{-}$}
 
The E2 transitions from the low-lying $3^{-}_{1}$ state to the dipole states, shown in Fig.\ref{sf_multi_xt1}(b),
 exhibits behavior different from that of Fig.\ref{sf_multi_xt1}(a).
The low-lying $3^{-}_{1}$ state has the collective character of the surface octupole vibration, including
 many particle-hole configurations of not only neutrons but also protons, as seen in the RPA amplitudes (Table \ref{tableXamp3}). 
 Here relevant to the E2 transition  are  proton
 particle-hole configurations in $3^{-}_{1}$ 
 since we use the bare charge for the E2 operator. 
 It is seen in Fig. \ref{sf_multi_xt1}(b) that there is no visible strength in the region
 of the soft dipole transition ($ S_{1n}  < E \lesssim 7$ MeV), and this is due to the neutron character of the
 soft dipole excitation. It is seen also that
 there exists several peaks in the energy region $8 \lesssim E \lesssim 17$ MeV, in contrast to
 the E1 transitions from the $2^{+}_{1,2}$. This originates from the
 relatively large number of proton particle-hole configurations mixed in the collective $3^{-}_{1}$ state.

\begin{table}
	\centering
	\caption{The RPA forward amplitudes $X_{ph}$ of the  $3^{-}_{1}$ state. Particle-hole configurations with large amplitude
	$|X_{ph}| > 0.1$ are listed. The neutron single-particle orbit $1i_{13/2}$ is a resonance in the continuum.}
    		\begin{tabular}[t]{ c c p{5mm} c c} \hline
		      	neutron config. &  $X^{3^{-}_{1}}_{ph}$ & &  proton config. &  $X^{3^{-}_{1}}_{ph}$ \\ \hline 
			$(1i_{13/2})(2f_{7/2})^{-1}$ & 0.831 && $(1h_{11/2})(1g_{9/2})^{-1}$ & -0.285  \\ 
			$(1i_{13/2})(1h_{11/2})^{-1}$ & 0.354 && $(1g_{7/2})(2p_{1/2})^{-1}$ & 0.203  \\ 
			$(1h_{9/2})(2d_{3/2})^{-1}$ & -0.299 && $(2d_{5/2})(2p_{1/2})^{-1}$ & 0.176  \\ 
			$(1h_{9/2})(1g_{7/2})^{-1}$ & -0.189 && $(2d_{5/2})(2p_{3/2})^{-1}$ & 0.135  \\
			$(2f_{5/2})(3s_{1/2})^{-1}$ & 0.134 && $(1j15/2)(1g_{9/2})^{-1}$ & 0.129  \\ 
			$(2g_{9/2})(2f_{7/2})^{-1}$ & 0.133 && $(2f_{7/2})(1g_{9/2})^{-1}$ & -0.129  \\ 
			$(2f_{5/2})(2d_{3/2})^{-1}$ & -0.126 && $(2d_{3/2})(2p_{3/2})^{-1}$ & -0.120  \\ 
			$(3p_{3/2})(2d_{3/2})^{-1}$ & -0.112 && $(3p_{3/2})(1g_{9/2})^{-1}$ & -0.103  \\ 
			$(2j15/2)(1g_{9/2})^{-1}$ & 0.108 && $(1g_{7/2})(1f_{5/2})^{-1}$ & 0.102  \\ 
			$(2f_{5/2})(1g_{7/2})^{-1}$ & -0.101 && & \\ \hline
   		 \end{tabular}
	\label{tableXamp3}
\end{table}   
 
 \subsubsection{Decay branching ratio from $1^{-}$ states}
 
\begin{figure}
	\centering
	\includegraphics[width=0.6\columnwidth]{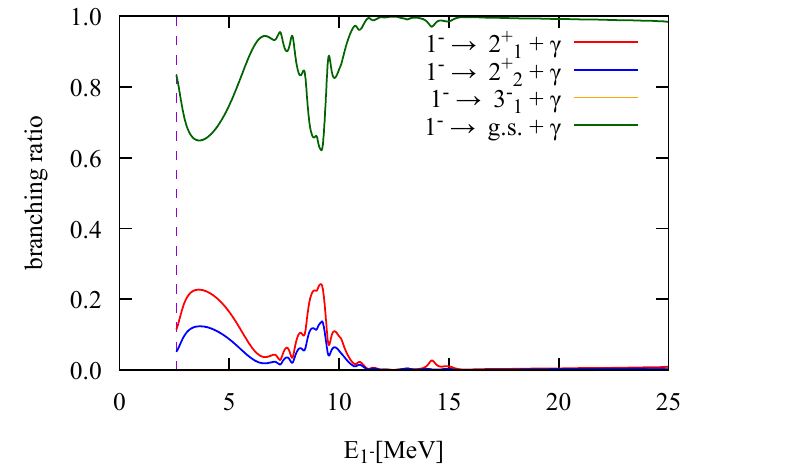}
	\caption{The branching ratios of photo-emission decays from the excited  $1^{-}$ states to  the ground state (green curve), the $2^{+}_{1}$ state (red curve) and the $2^{+}_{2}$ state (blue curve)
        in $^{140}{\rm Sn}$. Horizontal axis is the excitation energy of the $1^{-}$ states. The neutron separation energy $S_{1n}=2.59$ MeV 
        is indicated by dotted line. The branching ratio to the $3^{-}_{1}$ state is invisibly small.}
	\label{pe_1t2}
\end{figure}

 Combining the above results, we shall discuss the branching ratio for the photo-emission decays from
 excited $1^{-}$ states to the ground state, $2^{+}_{1,2}$  and
 $3^{-}_{1}$ states.
 The result is shown in Fig.\ref{pe_1t2}(a). It is seen that the soft dipole states 
 in the energy region $S_{1n} < E \lesssim 7$ MeV decays 
 not only to the ground state but also to both $2^{+}_{1}$ and $2^{+}_{2}$ states with sizable
 branching ratio $20-40\%$ (summing  $2^{+}_{1}$ and $2^{+}_{2}$). 
 This reflects that the neutron single-particle transition  
 $ (\mathrm{cont.}d_{5/2}) \to (3p_{3/2}) $ relevant to $1^{-} \leftrightarrow 2_{1,2}^{+}$ 
 is comparable to transition
 $(\mathrm{cont.}d_{5/2}) \to (2f_{7/2}) $ relevant to $1^{-} \leftrightarrow 0_{\mathrm{g}}^{+}$. 
  In the GDR region ($E \approx 10-17$ MeV), in contrast, the decay to the
 ground state is dominant because the lack of the E1 transition strengths to
 the $2^{+}_{1}$ and $2^{+}_{2}$. Note that the E2 decay probability
 to the $3^{-}_{1}$ state is negligibly small ($< 0.01\%$)  and not visible in the scale of Fig.\ref{pe_1t2}.
 
 \subsection{$2^{+}$ states: low-lying states and GQRs}
 
\begin{figure}
	\centering
	\includegraphics[width=0.6\columnwidth]{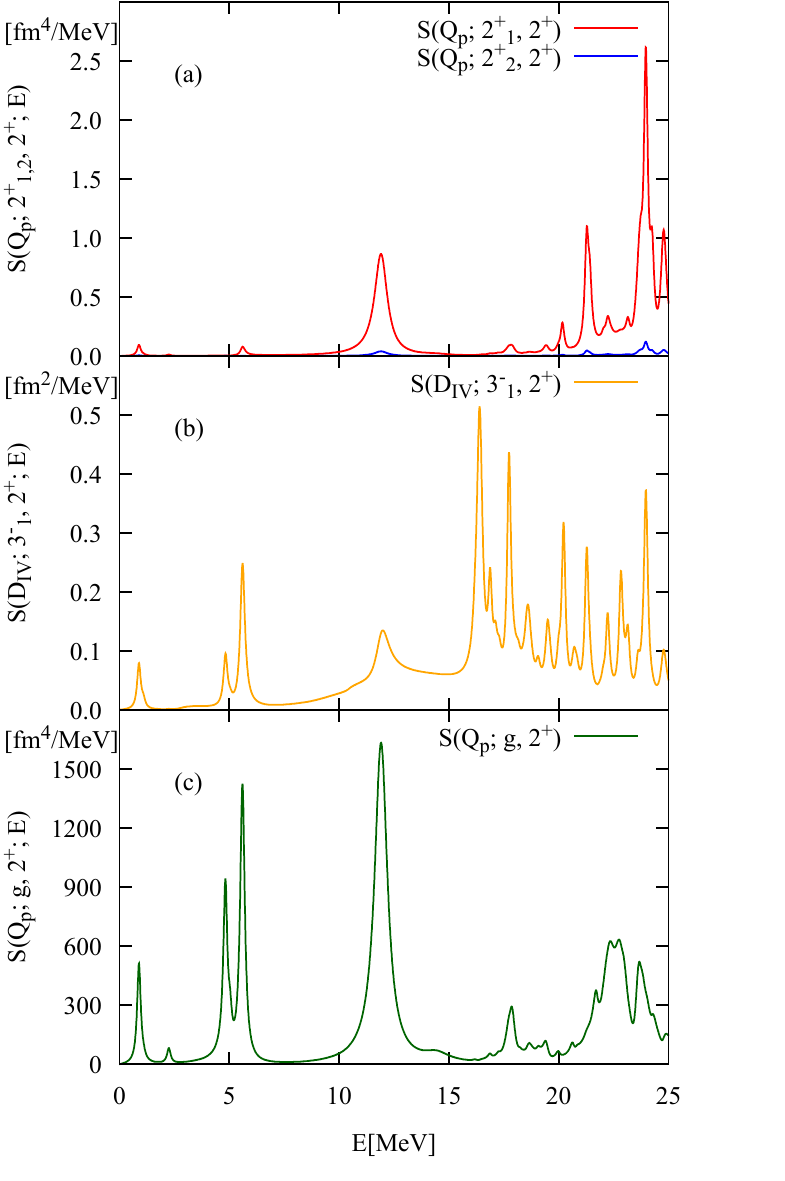}
	\caption{(a)The E2 strength functions $S(Q_{\mathrm{p}}; 2^{+}_{1,2}, 2^{+}; E)$ for transitions from $2^{+}_{1,2}$ to  $2^{+}$ states.
	(b) The E1 strength function $S(D_{\mathrm{IV}}; 3^{-}_{1}, 2^{+}; E)$ for transitions from $3^{-}_{1}$ to  $2^{+}$ states.
	(c) The E2 strength function $S(Q_{\mathrm{p}}; \mathrm{g}, 2^{+}; E)$ for 
	 transitions from the ground state to $2^{+}$ states. The horizontal axis is the excitation energy of the $2^{+}$ states.}
	\label{sf_multi_xt2}
\end{figure}

\begin{figure}
	\centering
	\includegraphics[width=0.6\columnwidth]{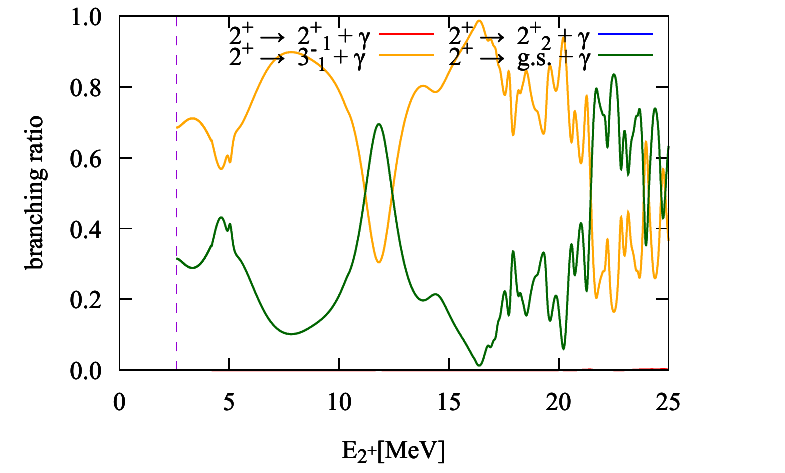}
	\caption{
	The branching ratios of photo-emission decays from the excited  $2^{+}$ states to 
	the low-lying $3^{-}_{1}$ state (yellow curve) and  the ground state (green one)
	in $^{140}{\rm Sn}$. The strengths associated with the low-lying $2^{+}$ states
	below the neutron separation energy (dotted line) are removed.
	Horizontal axis is the excitation energy of the $2^{+}$ states.
	The branching ratio to the $2^{+}_{1,2}$ states are invisibly small.
}
	\label{pe_2t3}
\end{figure}

Here we discuss excited $2^{+}$ states with focus on the GQR's and the low-lying $2^{+}$ states.

Figure \ref{sf_multi_xt2}(b)  shows 
the strength function $S(D_{\mathrm{IV}}; 3^{-}_{1}, 2^{+}; E)$ for the E1 transitions from $3^{-}_{1}$
to the excited $2^{+}$ states.
A 
peak around $E\approx 12$ MeV corresponds to
the transition from the low-lying collective $3^{-}_{1}$ state to  ISGQR.
We also observe another small peak at
$E \approx1 {\rm \, MeV}$. This is the transitions between the low-lying $2^{+}_{1,2}$ states and 
the collective $3^{-}_{1}$ state. (Note that the two $2^{+}$ states are not resolved due to the smoothing
with $\eta = 0.1 {\rm MeV}$.) These transitions can be described only if 
the correlation and the collectivity are taken into account in the theory.
Several peaks
at high energy region $E \gtrsim 16$ MeV and $E\sim 5$ MeV correspond to
 particle-hole configurations of both neutrons and protons, and existence of these transitions can be
 understood in terms of the same argument as that for the E2 transitions from $3^{-}_{1}$ to $1^{-}$ states.

The roles of the correlation and the collectivity are also seen
in the E2 transitions between $2^{+}_{1,2}$ and higher-lying $2^{+}$ states,  shown in
the strength function $S(Q_{\mathrm{p}}; 2^{+}_{1,2}, 2^{+}; E)$ (Fig. \ref{sf_multi_xt2}(a) ).
An example is 
the transition between  $2^{+}_{1}$ and ISGQR. This transition strength appears only if
configuration mixing of the proton particle-hole components is taken into account in
 the $2^{+}_{1}$ state.  
Note however that  the overall strengths 
in $S(Q_{\mathrm{p}}; 2^{+}_{1,2}, 2^{+}; E)$ 
between $2^{+}_{1,2}$ and higher-lying $2^{+}$ states (panel (a))
are significantly smaller than 
the E2 transition strengths from the ground state ($S(Q_{\mathrm{p}}; \mathrm{g}, 2^{+}; E)$ shown
panel (c)) due to the small admixture of proton configurations. 
The  $2^{+}_{2}$ state has even smaller admixture, as is suggested by the very small $B(E2, gs \to 2^{+}_{2})$
(see the inset of Fig.\ref{sf_gs}(b)), resulting in much smaller strengths than that of $2_{1}^{+}$.

Figure \ref{pe_2t3} shows the branching ratio of the photo-emission decays from the excited $2^{+}$ states to the
ground state (E2) and  the $3^{-}_{1}$ state (E1). The branching ratios of the E2 decays to the $2^{+}_{1,2}$ states 
is not shown here since they are negligibly small.  A gross behavior is that the E1 decay probability to $3^{-}_{1}$ state 
is larger than the E2 decays to the ground state in most of the plotted energy range except 
in the isoscalar and isovector GQR regions ($E \sim 12$ MeV and $E\sim 22-25$ MeV), In these
two energy regions, the collectivity of the GQR's enhances the E2 transition probability to the ground state and hence
the branching ratio. The collectivity of the ISGDR causes also enhancement
the transition to the $3^{-}_{1}$ state, but to a smaller extent than that to the ground state. 

\subsection{$3^{-}$ states: continuum, low-lying and high-lying collective states}

Concerning 
the excited $3^{-}$ states, we observe additional new features as well as similar behaviours to those found in the above examples.

\begin{figure}
	\centering
	\includegraphics[width=0.6\columnwidth]{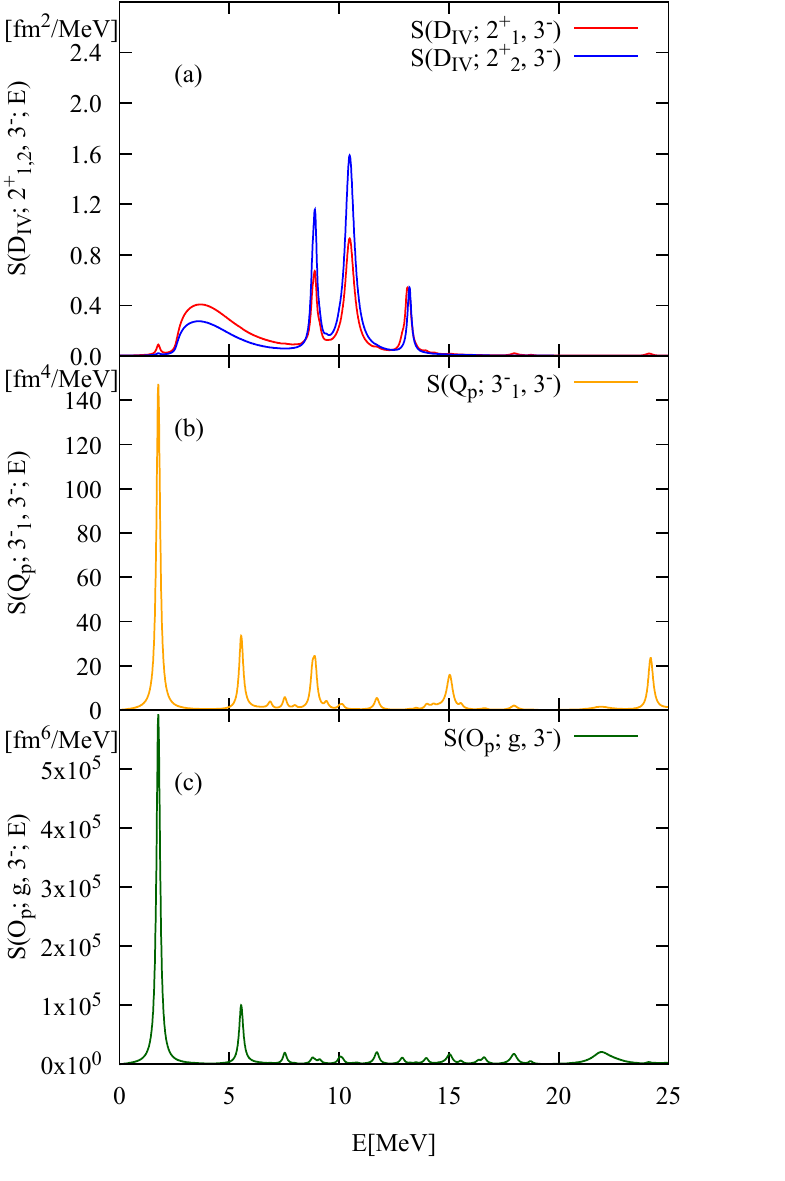}
	\caption{(a)The E1 strength functions $S(D_{\mathrm{IV}}; 2^{+}_{1,2}, 3^{-}; E)$ for transitions from $2^{+}_{1,2}$ to  $3^{-}$ states.
	(b) The E2 strength function $S(Q_{\mathrm{p}}; 3^{-}_{1}, 3^{-}; E)$ for transitions from $3^{-}_{1}$ to  $3^{-}$ states.
	(c) The E3 strength function $S(O_{\mathrm{p}}; \mathrm{g}, 3^{-}; E)$ for 
	 transitions from the ground state to $3^{-}$ states. The horizontal axis is the excitation energy of the $3^{-}$ states.}
	\label{sf_multi_xt3}
\end{figure}

\begin{figure}
	\centering
	\includegraphics[width=0.6\columnwidth]{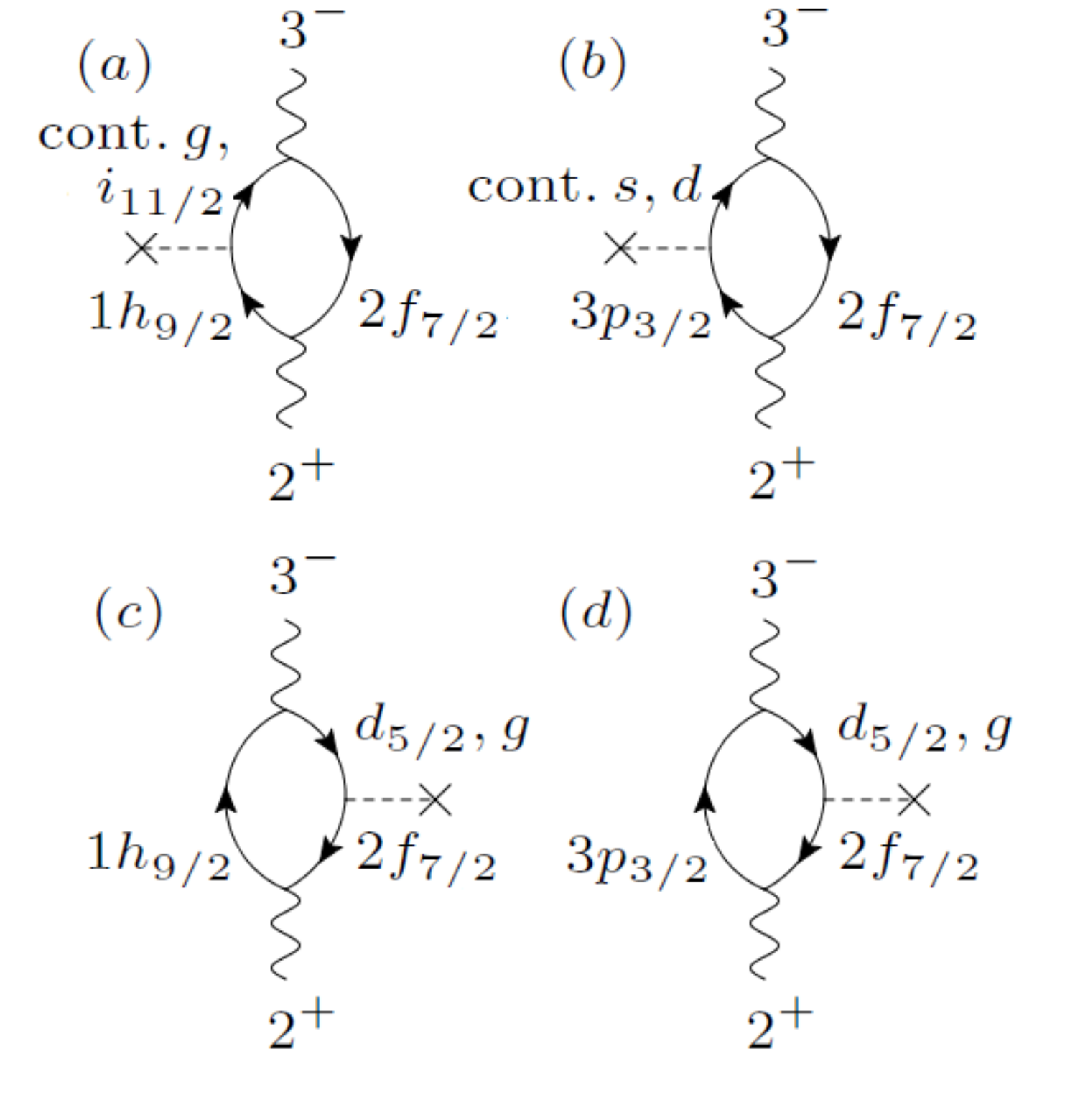}
	\caption{
	The diagrams representing dominant components of transition between the exited $3^{-}$  states and the low-lying $2^{+}_{1,2}$  in $^{140}{\rm Sn}$.
}
	\label{diagram_2t3}
\end{figure}

\begin{figure}
	\centering
	\includegraphics[width=0.6\columnwidth]{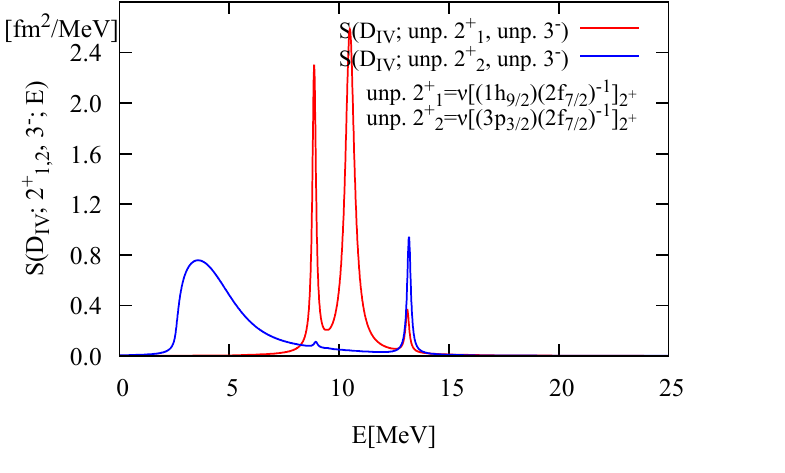}
	\caption{
	The E1 strength functions 
	$S(D_{\mathrm{IV}}; \nu [(1h_{9/2})(2f_{7/2})^{-1}]_{2^+}, 3^{-}; E)$ and $S(D_{\mathrm{IV}}; \nu [(3p_{3/2})(2f_{7/2})^{-1}]_{2^+}, 3^{-}; E)$
	for transitions from the neutron 1p-1h states
	$\nu [(1h_{9/2})(2f_{7/2})^{-1}]_{2^+}$ and 
	$\nu [(3p_{3/2})(2f_{7/2})^{-1}]_{2^+}$ to unperturbed $3^{-}$ states.
	The horizontal axis is the excitation energy of the $3^{-}$ states.
	}
	\label{sf_multi_xt3_nocorr}
\end{figure}

Figure \ref{sf_multi_xt3}(a) shows the strength function
$S(D_{\mathrm{IV}}; 2^{+}_{1,2}, 3^{-}; E)$ for the E1 transition from the low-lying $2^{+}_{1,2}$ to the $3^{-}$ states. 
A characteristic feature is  continuum strength in the region $2.59 (=S_{1n}) < E \lesssim 7$ MeV, as
is similarly seen in 
the strength function for the E1 transition $2^{+}_{1,2} \to 1^{-}$ (Fig.\ref{sf_multi_xt1}(a)).
Indeed this feature  can be understood 
in terms of the same argument using two dominant neutron particle-hole configurations 
$\nu[(1h_{9/2})(2f_{7/2})^{-1}]_{2^{+}}$ and $\nu[(3p_{3/2})(2f_{7/2})^{-1}]_{2^{+}}$
in the low-lying $2^{+}_{1,2}$ states.  Comparing with  
 the unperturbed transitions from these two configurations (Fig.\ref{sf_multi_xt3_nocorr}),  we find that 
the continuum strength is associated with 
continuum particle-hole states $\nu [(\mathrm{cont.}s_{1/2})(2f_{7/2})^{-1}]_{3^{-}}$
and $\nu [(\mathrm{cont.}d_{5/2,3/2})(2f_{7/2})^{-1}]_{3^{-}}$, which are excited 
from the configuration $\nu [(3p_{3/2})(2f_{7/2})^{-1}]_{2^{+}}$ by
neutron single-particle transition from $3p_{3/2}$  to continuum $s_{1/2}$ and  $d$ orbits,
(cf.  the diagram of Fig.\ref{diagram_2t3}(b)).
Other components shown in the diagrams Fig.\ref{diagram_2t3}(a)(c) and (d) brings 
three narrow peaks appearing around
$E\approx 9 - 13$  MeV.

We emphasize that the E1 transition between 
 the low-lying $2^{+}_{1,2}$ and the continuum octupole state is different from that between
 $2^{+}_{1,2}$ and the continuum dipole state: The strength of the former rises sharply
 at the threshold energy $E=2.59$ MeV ($=S_{1n}$). This originates from the transition
 $3p_{3/2}\to \mathrm{cont.}s_{1/2}$, where
the continuum s-orbit causes a cusp behavior at the threshold.
For $1^{-}$, however, the configuration
$\nu (\mathrm{cont.}s_{1/2})(2f_{7/2})^{-1}$ with the continuum $s$-orbit is forbidden by the angular momentum coupling.


Examples showing the collective effect are remarked also.
A small peak at $E \approx1.8$ MeV (below the neutron threshold energy)
corresponds to the E1 transition between the
low-lying octupole vibrational state $3^{-}_{1}$ at $E=1.77$ MeV and
 the low-lying $2^{+}_{1,2}$ states. Existence of the low-lying
collective state is a peculiar aspect of the $3^{-}$ channel.

Another example of the collective effect is seen in
Fig. \ref{sf_multi_xt3}(b), which shows the strength function
$S(Q_{\mathrm{p}}; 3^{-}_{1}, 3^{-}; E)$ of the E2 transitions between $3^{-}_{1}$ and
all the RPA excited states with $3^{-}$.  A 
peak at $E= 1.77$ MeV, which corresponds to the diagonal matrix element
$\bra{3^{-}_{1}} | \mathrm{E2} | \ket{3^{-}_{1}}$ for the low-lying collective state,
is significantly enhanced in comparison with the E2 matrix elements
associated with other non-collective $3^{-}$ states. The enhancement is caused by
the  collective and  surface vibrational character 
of the $3^{-}_{1}$ state.

\begin{figure}
	\centering
	\includegraphics[width=0.6\columnwidth]{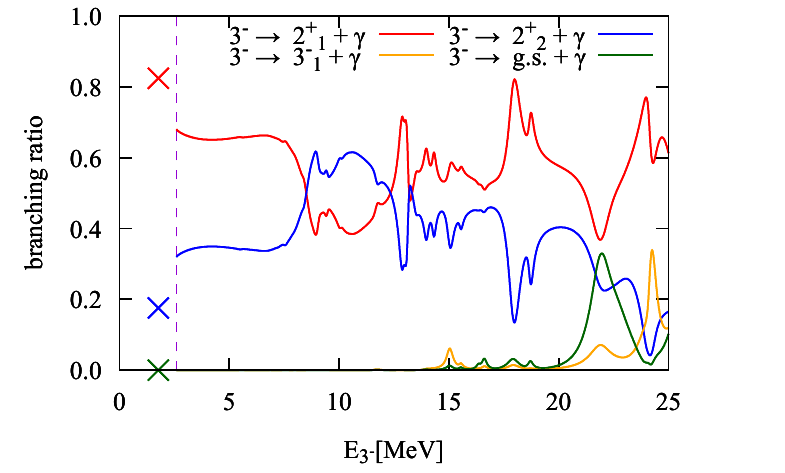}
	\caption{
	The branching ratios of photo-emission decays from the excited  $3^{-}$ states to  the ground state (green curve), the $2^{+}_{1}$ state (red curve), the $2^{+}_{2}$ state (blue curve) and the $3^{-}_{1}$ state (yellow)
	in $^{140}{\rm Sn}$. 
	Horizontal axis is the excitation energy of the $3^{-}$ states. The neutron separation energy is 2.59 MeV (dotted line).}
	\label{pe_3t2}
\end{figure}

Figure \ref{pe_3t2} shows the branching ratio of the photo-emission decays from $3^{-}$ states to  the
ground state (E3 transition) , the $2^{+}_{1,2}$ states (E1) and the $3^{-}_{1}$ state (E2).
The E1 transitions feeding to the low-lying   $2^{+}_{1,2}$ states dominate over the
E3 transition to the ground states in all the energy range from the continuum 
octupole states to the highest energy region
$E\sim 20-25$MeV. Looking at more details, the  branching ratios 
to the ground state,  $2^{+}_{1}$, $2^{+}_{2}$, and $3^{-}_{1}$ states reflect
various structures of the initial $3^{-}$ states. For example,
the branching ratios for decays from
continuum $3^{-}$ state around $2.59 < E \lesssim 7$ MeV to the two low-lying
$2^{+}_{1}$ and  $2^{+}_{2}$ states  are well accounted for by the mixing amplitudes
of the configuration $\nu [(2p_{3/2})(2f_{7/2})^{-1}]_{2^{+}}$ in $2^{+}_{1}$ and  $2^{+}_{2}$ (cf. Table.\ref{tableXamp2}).
 This indicates that the continuum octupole states in this
energy region is uncorrelated particle-hole excitations
$\nu [(\mathrm{cont.}s_{1/2})(2f_{7/2})^{-1}]_{3^{-}}$ and $\nu [(\mathrm{cont.}d_{3/2},1/2)(2f_{7/2})^{-1}]_{3^{-}}$.
The branching ratio
for transitions from the vibrational collective $3^{-}_{1}$ state to $2^{+}_{1}$ and  $2^{+}_{2}$
(the crosses at $E=1.77$ MeV) are different from that of the continuum states,
reflecting significant configuration mixing in the $3^{-}_{1}$ state (Table.\ref{tableXamp3}). Around $E \sim 22$ MeV, 
collectivity of the high-lying octupole vibrational state enhances the branching to the ground state.

\section{Conclusion}

The continuum random phase approximation (cRPA), referred also to the linear response theory,  describes both 
collective and
non-collective particle-hole excitations as well as their coupling to unbound single-particle orbits which
plays  key roles in exotic nuclei close to the proton and neutron drip-lines.  On the basis of the
nuclear density functional theory or the self-consistent Hartree-Fock model, cRPA
provides a well-defined scheme to calculate the response function for one-body field, e.g. the
electromagnetic matrix elements, for transitions from the ground state to the excited states. In the present study,
we have extended the linear response theory so that one can calculate transitions 
from a low-lying excited state to the RPA excited states  under consideration.
This extension enables us to calculate gamma-decays 
 from the RPA excited states to a set of the low-lying excited states, and hence
 the branching ratio and  the total decay probability including different final states. 
 
In order to demonstrate the applicability of the extended cRPA we have
 described the $1^{-}$, $2^{+}$ and
$3^{-}$ excitations  in a neutron-rich nucleus $^{140}$Sn. Specifically we discuss
the E1, E2 and E3 transitions among
low-lying vibrational states, higher-lying giant resonances and the unbound particle-hole states with
continuum spectra. 
An important conclusion is that there exist 
cases where the branching ratios  to the low-lying excited states are larger than or comparable with
 that  to the ground state. Furthermore
we have demonstrated that the extended cRPA enables us to
analyze  microscopic mechanisms how the transitions between the excited states reflect
the nature of the correlations in both the initial and final states. 

Let us remark future developments of the present study.  
Firstly, we plan to describe the radiative neutron-capture reaction 
of neutron-rich nuclei by applying the extended cRPA. In a preceding work\cite{Matsuo2015} we have 
formulated the theory of the direct neutron-capture reaction in which the 
reaction proceeds via the RPA correlated states, but we described  only the 
limited process where the final state
of the gamma-decays from the RPA states is  the ground state of the synthesized nucleus. Applying 
the extended cRPA, we can include decay channels populating the low-lying excited states, and hence
 provides more realistic description of the radiative neutron-capture reactions.
 Secondly, we plan to take into account the pair correlation which need to be included
 when the theory is applied to open-shell nuclei. Following
  Ref.\cite{Matsuo2001}, this can be achieved by replacing the RPA with the
  quasiparticle random phase approximation (QRPA).
  
\section{Acknowledgments}
The authors thank Kazuyuki Sekizawa for valuable discussions. This work was
supported by the JSPS KAKENHI (Grant No. 20K03945).

\appendix

\section{Pseudo transition density-matrix in the linear response formalism}

The pseudo transition density-matrix can be calculated as follows.
Suppose that  we describe the low-lying excited state $\ket{i}$ using 
the linear response equation
\begin{align}
	\delta \rho^{'}(\mathbf{r}_{x}, \omega) 
	= \sum_{\sigma_{x}} \int dx^{'} R_{0}(x, x; x^{'}, x^{'}; \omega) \frac{\delta U}{\delta \rho}(\mathbf{r}_{x^{'}}) 
	\delta \rho^{'}(\mathbf{r}_{x^{'}}, \omega) + 
\sum_{\sigma_{x}} \int dx^{'} R_{0}(x, x; x^{'}, x^{'}; \omega) {f}^{'}(\mathbf{r}_{x^{'}}) 
\label{delta_rho_i}
\end{align}
and an external perturbation $\hat{M}^{'}=\int dx {f}^{'}(\mathbf{r}_{x}) \hat{\rho}(x)$, which is suitable to excite $\ket{i}$.
(Equation (\ref{delta_rho_i}) is essentially the same as  Eq.(\ref{local_linear_response_eq}) except the difference in the external perturbation. 
We put prime ${}^{'}$ to the density response  $\delta \rho^{'}(\mathbf{r}_{x},\omega)$ in order to distinguish it from 
$\delta \rho(\mathbf{r}_{x},\omega)$ in Eq.(\ref{local_linear_response_eq}).) 
We can also consider the extended linear response equation for the density-matrix  response $\delta \rho^{'}(x,y,\omega)$
in terms of an equation similar to Eq.(\ref{general_linear_response_eq}).

The low-lying RPA  state $\ket{i}$ 
appears as a pole at $\omega_i=(E_i-E_0)/\hbar$,
 in $\delta \rho^{'}(\mathbf{r}_{x},\omega)$ provided that $\ket{i}$ 
is a discrete bound state. 
The transition density $\rho^{(\mathrm{tr})}_{ i}(\mathbf{r}_{x},\omega) \equiv \sum_{\sigma_{x}}\bra{0} \hat{\rho}(x) \ket{i}$
corresponds to the residue of  $\delta \rho^{'}(\mathbf{r}_{x},\omega)$ at the pole, and thus
can be calculated with
\begin{align}
\rho^{(\mathrm{tr})}_{ i}(\mathbf{r}_{x},\omega) = C' \mathrm{Im} \delta \rho^{'}(\mathbf{r}_{x},\omega_i) 
\end{align}
where  $C'$ is a normalization constant. 
Similarly the transition density-matrix $\rho^{(\mathrm{tr})}_{ i}(x, y)$ 
is also given by
\begin{align}
	\rho^{(\mathrm{tr})}_{i}(x,y) &=C' \mathrm{Im}\delta \rho^{'}(x, y, \omega_{i})  \notag \\
	& = C' \mathrm{Im} 
	\left\{ \int dx^{'} R_{0}(x, y; x^{'}, x^{'}; \omega_{i}) \frac{\delta U}{\delta \rho} (\mathbf{r}_{x^{'}}) 	
		        \delta \rho^{'}(\mathbf{r}_{x},\omega_i) 
		    +    \int dx^{'} R_{0}(x, y; x^{'}, x^{'}; \omega_{i}) {f}^{'}(\mathbf{r}_{x^{'}})  \right\}.
\end{align}

As we discussed for Eq.(\ref{pseudo_density_matrix}), the  pseudo transition density-matrix $\bar{\rho}^{(\mathrm{tr})}_{i}(x, y)$ 
has the same structure as that of the transition density-matrix except the sign of the backward amplitudes.
Thus it is calculated with
\begin{align}
	 \bar{\rho}^{(\mathrm{tr})}_{ i}(x, y) = C'  \mathrm{Im}
	\left\{  \int dx^{'} \bar{R}_{0}(x, y; x^{'}, x^{'}; \omega_{i}) \frac{\delta U}{\delta \rho} (\mathbf{r}_{x^{'}}) 
	 		       \delta \rho^{'}(\mathbf{r}_{x},\omega_i)
			 +    \int dx^{'} \bar{R}_{0}(x, y; x^{'}, x^{'}; \omega_{i}) {f}^{'}(\mathbf{r}_{x^{'}})  \right\},
\label{bar_delta_rho}
\end{align}
\begin{align}
	\bar{R}_{0}(x, y; y^{'}, x^{'}; \omega) \equiv \sum_{h} &\left\{ \phi^{*}_{h}(y)\bar{G}_{0}(x, x^{'}, \epsilon_{h} + \hbar \omega + i\eta)\phi_{h}(y^{'}) \right. \notag \\ 
	&\left. - \phi^{*}_{h}(x^{'})\bar{G}_{0}(y^{'}, y, \epsilon_{h} - \hbar \omega - i\eta)\phi_{h}(x) \right\},
\end{align}
where the function $\bar{R}_0$  is a variant of the unperturbed response function $R_0$ with
 the  sign of the second term opposite to that of  Eq.(\ref{unp_respfn_G}). 
Note that the Green's function $G_{0}$ is replaced with
\begin{align}
	\bar{G}_{0}(x, x^{'}, e) \equiv G_{0}(x, x^{'}, e) - \sum_{h} \frac{\phi_{h}(x) \phi^{*}_{h}(x^{'})}{e - \epsilon_{h}}.
\end{align}
so that the contribution of the hole orbits in the Green's function are removed. This replacement is necessary 
to remove hole-hole components in $\bar{R}_{0}$, which are automatically canceled out in the original unperturbed response function $R_{0}$.

\section{Response functions for spherical mean-field}

Assuming the spherical symmetry of the mean-field, we represent
the single-particle wave function by
$
	\phi_{nljm}(x) = Y_{ljm}(\hat{x}) \frac{1}{r_{x}} \phi_{nlj}(r_{x}),
$
where $r_{x}$ and $\hat{x} \equiv (\hat{\mathbf{r}}_{x}, \sigma_{x})$ are the radial and angle-spin variables, respectively and
$Y_{ljm}(\hat{x})$ is the spin spherical harmonics with
the angular quantum numbers $ljm$.

The single-particle Green's function is given by
\begin{align}
	G_{0}(x, x^{'}, E) = \sum_{ljm} Y_{ljm}(\hat{x}) \frac{1}{r_{x} r_{x^{'}}} G_{0, lj}(r_{x}, r_{x^{'}}, E) Y^{*}_{ljm}(\hat{x}^{'}).
\end{align}
whose radial part can be constructed exactly as
\begin{align}
	G_{0, lj}(r_{x}, r^{'}_{x}, E) = \frac{2m}{\hbar^{2}} \frac{1}{W(\phi_{1, lj}, \phi_{2, lj})} \left\{ \phi_{1, lj}(r^{'}_{x})\phi_{2, lj}(r_{x}) \theta(r_{x} - r^{'}_{x}) + \phi_{1, lj}(r_{x}) \phi_{2, lj}(r^{'}_{x}) \theta(r^{'}_{x} - r_{x}) \right\}
\end{align}
in terms of the regular radial wave $\phi_{1, lj}(r)$ and the outgoing wave $\phi_{2, lj}(r)$ with a given complex energy $E$. 
$W$ is the Wronskian.

 The unperturbed response function for density matrix and non-local one-body operators is represented by
\begin{align}
	R_{0}(x, y; y^{'}, x^{'}; \omega) = \sum_{ljm, l^{'}j^{'}m^{'}} Y_{l^{'}j^{'}m^{'}}(\hat{x}) Y^{*}_{ljm}(\hat{y}) \frac{1}{r_{x} r_{y} r_{y^{'}} r_{x^{'}}} R_{0, l^{'}j^{'}, lj}(r_{x}, r_{y}; r_{y^{'}}, r_{x^{'}}; \omega) Y_{ljm}(\hat{y}^{'}) Y^{*}_{l^{'}j^{'}m^{'}}(\hat{x}^{'}).
\end{align}
Here the radial unperturbed response function is given by
\begin{align}
	R_{0, l^{'}j^{'}, lj}(r_{x}, r_{y}; r_{y^{'}}, r_{x^{'}}; \omega) = \sum_{n} &\Bigl\{ \phi^{*}_{nlj}(r_{y}) G_{0, l^{'}j^{'}}(r_{x}, r_{x^{'}}, \epsilon_{nlj} + \hbar \omega + i \eta) \phi_{nlj}(r_{y^{'}}) \theta(\epsilon_{F} - \epsilon_{nlj}) \notag \\ 
&+ \phi^{*}_{nl^{'}j^{'}}(r_{x^{'}}) G_{0, lj}(r_{y^{'}}, r_{y}, \epsilon_{nl^{'}j^{'}} - \hbar \omega - i \eta) \phi_{nl^{'}j^{'}}(r_{x}) 		
	\theta(\epsilon_{F} - \epsilon_{nl^{'}j^{'}}) \Bigr\}.
\end{align}
where $\phi_{nlj}(r)$ is a radial wave function of the single-particle states occupied in the ground state. 

We define the creation operator of an RPA excited state and its forward and backward amplitudes by
\begin{align}
	\hat{O}^{\dag}_{i L_{i}M_{i}} &= \sum_{ph} \left\{ X^{i}_{ph} [a^{\dag}_{p} a_{h}]_{L_{i} M_{i}} - Y^{i}_{ph} [a^{\dag}_{h} a_{p}]_{L_{i} M_{i}} \right\}, \notag \\
	 [a^{\dag}_{p} a_{h}]_{L_{i} M_{i}} &=  \sum_{m_{p} m_{h}} 
	 \langle j_{p} m_{p} j_{h} m_{h} | L_{i}M_{i} \rangle a^{\dag}_{n_{p} l_{p} j_{p} m_{p}} a_{\widetilde{n_{h} l_{h} j_{h}m_{h}}},
\end{align}
where $a_{\widetilde{n l j m}}$ is the time reversal of the Fermion annihilation operator $a_{nljm}$.

\bibliography{extoex_cRPA_refs}

\end{document}